\documentclass[times,twocolumn,final]{elsarticle}

\usepackage{medima}
\usepackage{framed,multirow}
\usepackage{booktabs} 

\usepackage{amssymb}
\usepackage{latexsym}

\usepackage{url}
\usepackage{xcolor}

\usepackage{hyperref}
\usepackage{amsmath}

\definecolor{newcolor}{rgb}{.8,.349,.1}

\journal{Medical Image Analysis}

\begin{document}

\verso{Hao Zhang \textit{et~al.}}

\begin{frontmatter}

\title{Deep Unfolding Network with Spatial Alignment for Multi-modal MRI Reconstruction}%

\author[1]{Hao \snm{Zhang}}
\cortext[cor1]{Corresponding author:}
\author[1]{Qi \snm{Wang}}
\author[2]{Jun \snm{Shi}}
\author[3,4]{Shihui \snm{Ying}\corref{cor1}}
\ead{shying@shu.edu.cn}
\author[1]{Zhijie \snm{Wen}}

\address[1]{Department of Mathematics, School of Science, Shanghai University, Shanghai 200444, China}
\address[2]{School of Communication and Information Engineering, Shanghai University, Shanghai 200444, China}
\address[3]{Shanghai Institute of Applied Mathematics and Mechanics, Shanghai University, Shanghai 200072, China}
\address[4]{School of Mechanics and Engineering Science, Shanghai University, Shanghai 200072, China}

\begin{abstract}
Multi-modal Magnetic Resonance Imaging (MRI) offers complementary diagnostic information, but some modalities are limited by the long scanning time. To accelerate the whole acquisition process, MRI reconstruction of one modality from highly under-sampled k-space data with another fully-sampled reference modality is an efficient solution. However, the misalignment between modalities, which is common in clinic practice, can negatively affect reconstruction quality. Existing deep learning-based methods that account for inter-modality misalignment perform better, but still share two main common limitations: (1) The spatial alignment task is not adaptively integrated with the reconstruction process, resulting in insufficient complementarity between the two tasks; (2) the entire framework has weak interpretability. In this paper, we construct a novel Deep Unfolding Network with Spatial Alignment, termed DUN-SA, to appropriately embed the spatial alignment task into the reconstruction process. Concretely, we derive a novel joint alignment-reconstruction model with a specially designed aligned cross-modal prior term. By relaxing the model into cross-modal spatial alignment and multi-modal reconstruction tasks, we propose an effective algorithm to solve this model alternatively. Then, we unfold the iterative stages of the proposed algorithm and design corresponding network modules to build DUN-SA with interpretability. Through end-to-end training, we effectively compensate for spatial misalignment using only reconstruction loss, and utilize the progressively aligned reference modality to provide inter-modality prior to improve the reconstruction of the target modality. Comprehensive experiments on four real datasets demonstrate that our method exhibits superior reconstruction performance compared to state-of-the-art methods.
\end{abstract}


\end{frontmatter}


\section{Introduction}
\label{sec:introduction}
Magnetic Resonance Imaging (MRI), due to its non-invasive, high resolution, and significant soft tissue contrast, has become a widely used medical imaging technique. However, the MR scan is relatively slow due to the repetitive acquisition of MR signal spatial encoding and hardware limitations. This time-consuming process can lead to discomfort for patients, causing them to move and introducing motion artifacts in the images, which may negatively affect subsequent disease diagnosis. Therefore, accelerating MRI acquisition is of great importance in clinic practice. To accelerate MRI acquisition, a feasible strategy is to reduce the amount of k-space data collected and then reconstruct fully-sampled images by under-sampled data.

The Compressed Sensing MRI (CS-MRI) methods enable the accurate reconstruction from under-sampled data at sampling rates significantly below those required by the Nyquist sampling theorem \citep{https://doi.org/10.1002/mrm.21391}. They aim at designing some hand-crafted regularizers based on different priors (e.g., structured sparsity \citep{LAI201693,6862879}, non-local sparsity \citep{QU2014843,eksioglu2016decoupled}) of MR images and formulating them into the algorithm optimization to constrain the solution space. Regardless of its theoretical guarantees, it is challenging to handcraft an optimal regularizer. Alternatively, deep learning-based methods have attracted widespread attention in MRI reconstruction due to their accuracy and speed \citep{7493320,10.1007/978-3-030-59713-9_7,Zhu2018}. By learning robust feature representations, the deep learning-based methods have achieved impressive reconstruction performance. However, most deep learning-based methods are a black-box process and lack interpretability, which is required in clinical practice. To alleviate this black-box issue, deep unfolding networks \citep{8550778,pmlr-v172-xin22a,10318101,zhang2022high,jiang2023latent,jiang2023ga} have been proposed to incorporate the imaging model and domain knowledge into the network. They unfold the iterations of an optimization algorithm into deep neural networks, thus making the learning process interpretable.

Despite the potential of deep unfolding networks in MRI reconstruction, most of them focus on utilizing information from a single modality. However, in clinical practice, it is common to acquire MR images of different contrasts because each modality reveals distinct tissue and organ characteristics, and the complementary information among modalities contributes to more accurate diagnoses. While different modalities of MR images display different signal types, they are spatially corresponding and depict the same anatomical structures. Researches \citep{10.1007/978-3-030-00928-1_25, 9796552,bian2022learnable,ijcai2023p112} have shown that reconstruction of one modality (target modality) can be improved by utilizing the information from another modality (reference modality). But, these multi-modal MRI reconstruction methods are all based on the assumption that the images are perfectly aligned, which is rare in practice. The misalignment may negatively affect the reconstruction performance due to insufficiently exploring the correlation of different modalities.

To mitigate the misalignment between modalities, \cite{LAI201795} alternating iteration of the registration and the reconstruction to align the reference modality with the intermediate reconstruction result of the under-sampled target modality. However, the use of conventional iterative optimization is relatively time-consuming. Deep learning-based methods that take inter-modality misalignment into consideration perform better in terms of reconstruction and spatial alignment accuracy, but still share two main common limitations: (1) The spatial alignment is not adaptively integrated with the reconstruction process,  resulting in insufficient complementarity between the two tasks; (2) the entire framework has weak interpretability. For example, before reconstruction, \cite{9745968} and \cite{LIU202133} simply align the images of the reference modality with the under-sampled images of the target modality, neglecting the negative effects that artifacts in the under-sampled images can have on spatial alignment.

To alleviate the aforementioned limitations, we propose a Deep Unfolding Network with Spatial Alignment (DUN-SA) framework in this paper. Specifically, we first derive a novel joint alignment-reconstruction model, in which an aligned cross-modal prior term is developed to compensate for misalignment between modalities. Subsequently, we relax it into cross-modal spatial alignment and multi-modal reconstruction tasks, and propose an optimization algorithm for these two tasks. Concretely, we use the gradient-based algorithm for spatial alignment and the Half-Quadratic Splitting (HQS) algorithm for reconstruction. By alternately optimizing the two tasks, we solve this model. Further, we unfold the iterative stages of the proposed algorithm and design corresponding network modules. Finally, we propose an end-to-end, deep unfolding network with interpretability. The main contributions of this paper are as follows:
\begin{itemize}
    \item We propose a novel joint alignment-reconstruction model for multi-modal MRI reconstruction, in which an aligned cross-modal prior term is developed to compensate for misalignment between modalities while learning inter-modality prior. Utilizing gradient-based and HQS techniques, we design an optimization algorithm to alternately solve this model.

    \item By unfolding the iterative stages of the proposed algorithm and integrating them with specially designed network modules, we construct a deep unfolding network, termed DUN-SA, which exhibits clear interpretability.

    \item We design the Aligned Inter-modality Prior Learning Block (AIPLB) to learn inter-modality prior through aligned images from reference modality, and utilize the Denoising Block (DB) to fully exploit intra-modality prior.

    \item Through extensive experiments on the fastMRI dataset, the IXI dataset, the In-house dataset, and the BraTs 2018 dataset, we demonstrate that the proposed DUN-SA outperforms existing state-of-the-art methods in terms of quantitative and qualitative reconstruction results.
\end{itemize}

\section{Related Work}
\subsection{Single-Modal CS-MRI}
The single-modal MRI reconstruction problem is to reconstruct the original image from its partial acquisition. Traditionally, model-based methods usually utilize image prior to improve reconstruction performance. \cite{LAI201693,6862879} use wavelet transform to sparsely represent magnetic resonance images in iterative image reconstructions, capitalizing on sparse prior. \cite{QU2014843,eksioglu2016decoupled} further sparsify magnetic resonance images by exploiting the similarity of image patches, emphasizing non-local regularization. \cite{5617283,7337391} propose dictionary learning for adaptively learning the sparsifying transform (dictionary), and reconstructing the image simultaneously from highly under-sampled data.

Deep learning methods, with their powerful ability to represent features, can achieve superior reconstruction quality and acceleration compared to non-deep learning approaches. \cite{7493320} utilize CNN architectures to learn the mapping relationship between the MR images obtained from zero-filled and fully-sampled k-space data, significantly improving reconstruction speed while maintaining image quality. \cite{8233175,NEURIPS2020_567b8f5f} not only reconstruct images but also retain more image texture details and reduce more image artifacts through adversarial learning. Contrary to models that reconstruct under-sampled inputs in the image domain, \cite{10.1007/978-3-030-59713-9_7,Zhu2018} operate on under-sampled k-space, achieving better reconstruction performance. \cite{SHAUL2020101747,9084142,ZHANG201990,https://doi.org/10.1002/mrm.27201} utilize cross-domain networks for image reconstruction, leveraging information in both image domain and k-space, outperforming single-domain methods. Furthermore, \cite{WANG20201} incorporate the wavelet domain, using information across three domains, further improving reconstruction accuracy.

\subsection{Multi-Modal CS-MRI}
In multi-modal MRI reconstruction, a reference modality can guide the reconstruction of the target modality. Model-based methods explore the relationship between modalities based on prior knowledge. \cite{doi:10.1137/15M1047325} introduce two types of total variation based on position and direction, taking into account the structural prior of MR images. \cite{https://doi.org/10.1118/1.4962032} acknowledge the differences between modalities and proposes an iterative weighted reconstruction approach. \cite{8786180} present a method based on coupled dictionary learning for multi-contrast MRI reconstruction, effectively utilizing the structural dependencies between different contrasts. \cite{LAI201795} address the misalignment issue between modalities, proposing a novel MRI image reconstruction method that learns prior from multi-contrast images through a graph wavelet representation, modeling it as a bi-level optimization problem to allow misalignment between these images.

Deep learning-based methods \citep{8758456,8552399,9115255} combine T1-weighted and T2-weighted images as dual-channel inputs to a deep learning model, improving reconstruction quality by leveraging inter-modality complementary information. \citep{Zhou_2020_CVPR} integrate multi-modal and dual-domain information, further enhancing reconstruction accuracy. \cite{9796552} employ a Transformer structure, using multi-head attention mechanism to deeply capture multi-modal information, offering more global information compared to existing CNN-based methods. Methods that take misalignment between modalities into consideration perform better. \cite{LIU202133} utilize CNN to learn rigid transformation for compensating misalignment between unregistered paired multi-modal MR images, thereby making more efficient use of the reference modality. \cite{9745968} employ a spatial alignment network to compensate misalignment between modalities and incorporate a novel loss function that efficiently trains both the spatial alignment network and the reconstruction network simultaneously.

\subsection{Deep Unfolding Network}
Deep Unfolding Networks have achieved impressive results in many medical applications (e.g. dynamic MR imaging \citep{HUANG2021102190}, metal artifact
reduction \citep{WANG2023102729,10.1007/978-3-030-87231-1_11}, MRI Super-resolution \citep{YANG2023107605,10.1145/3503161.3548068,ijcai2023p112} and CS-MRI \citep{8550778,10.1007/978-3-030-59713-9_19,ijcai2023p112}), thereby attracting widespread attention recently. By unfolding certain optimization algorithms with network modules, They integrate model-based and learning-based methods well.

In single-modal CS-MRI, for example, by unfolding different optimization algorithms such as Alternating Direction Method of Multipliers (ADMM) \citep{8550778,10318101,jiang2023latent}, Alternating Iterative Shrinkage-thresholding Algorithm(ISTA) \citep{zhang2022high}, Half-Quadratic Splitting (HQS) \citep{pmlr-v172-xin22a,jiang2023ga} and combining them with different deep neural networks (e.g., CNNs, ResNets, U-nets), deep unfolding networks achieve sparsity in the denoising perspective. This approach not only improves reconstruction quality but also offers interpretability. Additionally, research has shown that incorporating multi-modal information can improve the quality of reconstruction, prompting the development of deep unfolding networks for multi-modal MRI reconstruction. \cite{10.1007/978-3-030-59713-9_19} capitalize on cross-modal prior and combine channel and spatial attention mechanisms to construct proximal operators, significantly enhancing image reconstruction accuracy. \cite{ijcai2023p112} employ Convolutional Sparse Coding (CSC) techniques for multi-modal MRI modeling. It transfers the common texture information from the reference modality images to the target modality images while avoiding the interference of inconsistent information and integrates deep network modules, achieving superior reconstruction results.

\section{Method}
\label{method}
In Section \ref{method}, we propose a joint alignment-reconstruction model for multi-modal MRI reconstruction and design an optimization algorithm in Section \ref{SAAMCSA}. Then, we unfold the iterative stages of the proposed algorithm with corresponding network modules and build DUN-SA in Section \ref{DUNWSA}. Finally, we introduce the details of network parameters and network training in Section \ref{NPLF}.
\subsection{Joint alignment-reconstruction model}
\label{SAAMCSA}
\subsubsection{The objective functions}
\label{of}
 Let \( x \in \mathbb{C}^{N} \) represent the fully-sampled MRI image, and \( k \in \mathbb{C}^N \) denote the measured fully-sampled frequency-domain signal with $N$ being the amount of pixels in the acquired image. Here, \(k\) is the Fourier transform of \(x\), directly measured in MRI data acquisition. In the classical compressed sensing problem for MRI, \( \tilde{k} \in \mathbb{C}^M \) is used to represent the under-sampled signal, expressed by applying a Fourier transform and a masking operator to the fully-sampled image. Consequently, the relationship between the fully-sampled MRI image and the under-sampled k-space signal can be formulated as:
\begin{equation}
\tilde{k} = F_{m}x + n,  \tag{1} \label{Eq:1}
\end{equation}
where \( F_m = MF \) represents masked Fourier transform, \( F \) is the Fourier transform operator, \( M \) is the masking operator, and \( n \) is the measurement noise generally considered as Gaussian noise. Additionally, \( M \ll N \), and \( \frac{M}{N} \) indicates the compression ratio. \( M \) is significantly smaller than \( N \), which leads to an underdetermined inverse problem. 

The classical compressed sensing method achieves reconstruction by optimizing the following energy function:
\begin{equation}
\mathop{\min}_{x} \frac{1}{2} \| F_m x - \tilde{k} \|^{2}_2 + \eta R(x),  \tag{2} \label{Eq:2}
\end{equation}
where the first term is the fidelity term, ensuring that the reconstructed image \( x \) is consistent with the under-sampled signal \( \tilde{k} \) in k-space; the second term \( R(x) \) is the regularization term, enforcing prior constraint like sparsity, and \( \eta \) is the balancing factor.

In multi-modal MRI reconstruction, an auxiliary term measures the correlation between \( x \) and \( x_{\text{ref}} \), where the reference modality provides extra inter-modality prior to assist the reconstruction of the target modality. The energy function becomes:
\begin{equation}
\mathop{\min}_{x} \frac{1}{2} \| F_m x - \tilde{k} \|^{2}_2 + \lambda \Psi(x, x_{\text{ref}}) + \eta R(x),  \tag{3} \label{Eq:3}
\end{equation}
where \( x_{\text{ref}} \) represents the fully-sampled image from the reference modality of the same region of interest, \(\Psi(x, x_{\text{ref}})\) models the correlation between \(x\) and \( x_{\text{ref}} \), and \( \lambda \) is the balancing coefficient.

Furthermore, to compensate for the spatial misalignment between \( x \) and \( x_{\text{ref}} \), we transform the auxiliary term into an aligned cross-modal prior term, integrating the spatial alignment task into the reconstruction process, yielding the final energy function:
\begin{equation}
\mathop{\min}_{x,\phi} \frac{1}{2} \left\|F_m x-\tilde{k}\right\|^2+\lambda \Psi(x,\mathcal{T}(x_{\text{ref}}, \phi))+\eta R(x),  \tag{4} \label{Eq:4}
\end{equation}
where \( \phi \) represents the displacement field, \( \mathcal{T} \) denotes the warp operation, and \( \mathcal{T}(x_{\text{ref}}, \phi) \) represents the aligned reference modality image. We refer to \( \Psi(x,\mathcal{T}(x_{\text{ref}}, \phi)) \) as the aligned cross-modal prior term, which models the correlation between target modality and aligned reference modality. In order to use standard gradient-based methods to optimize $\phi$, we use a differentiable warp operation \citep{jaderberg2015spatial}  to compute \(\mathcal{T}(x_{\text{ref}}, \phi)\). For each position \( p \), we compute a new position \( p' = p + \phi(p) \) in \( x_{\text{ref}} \) resulting in the following expression \(\mathcal{T}(x_{\text{ref}},\phi)[p] = x_{\text{ref}}[p + \phi[p]]\). Bilinear interpolation is used to sample from non-integer positions, ensuring the smoothness of the transformation.

We set Eq. \eqref{Eq:4} as our final objective function. By eliminating misalignment between modalities, we can utilize the information between corresponding points in aligned reference modality and target modality images to learn inter-modality prior more efficiently, and further improve the reconstruction results.

\subsubsection{The optimization algorithm}
\label{toa}
\begin{figure*}[!t]
\centerline{\includegraphics[width=\textwidth]{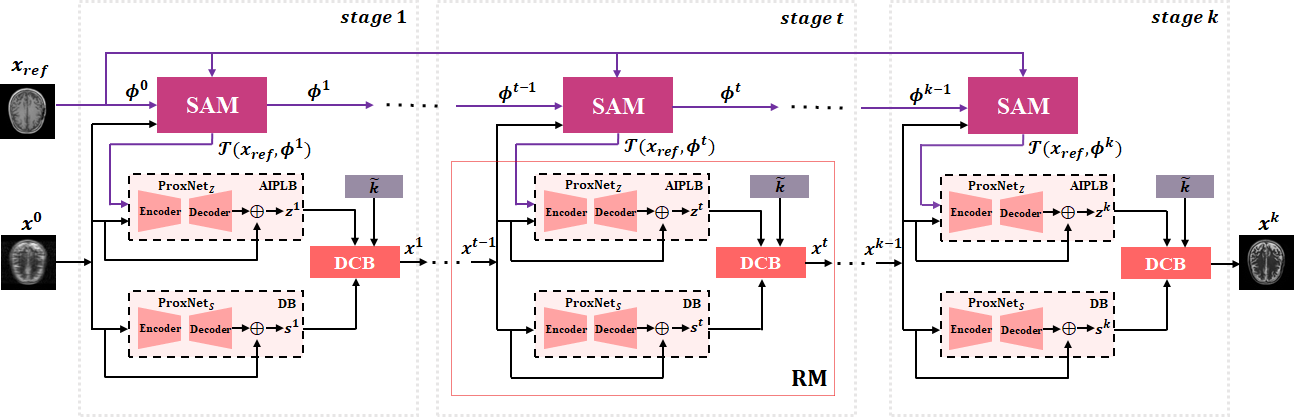}}
\caption{The overall structure of the proposed Deep Unfolding Network with Spatial Alignment (DUN-SA) consists of SAM (Spatial Alignment Module) and RM (Reconstruction Module). The RM is composed of AIPLB (Aligned Inter-modality Prior Learning Block), DB (Denoising Block), and DCB (Data Consistency Block). SAM is used to solve spatial alignment task, while RM is for reconstruction task. Specifically, AIPLB is used to learn aligned inter-modality prior, DB is used to learn denoising prior, and DCB is used to enforce data consistency constraint.}
\label{fig1}
\end{figure*}
To solve this model effectively, we relax Eq. \eqref{Eq:4} into cross-modal spatial alignment and multi-modal reconstruction tasks and optimize them alternatively:
\begin{align}
\hat{\phi} &= \mathop{\arg\min}_{\phi} \lambda \Psi(x,\mathcal{T}(x_{\text{ref}}, \phi)), \tag{5a} \label{eq:5a}\\
\hat{x} &= \mathop{\arg\min}_{x} \frac{1}{2} \left\|F_m x-\tilde{k}\right\|^2+\lambda \Psi(x,\mathcal{T}(x_{\text{ref}}, \phi)) +\eta R({x}). \tag{5b}\label{eq:5b}
\end{align}

\textbf{Update \(\phi\):} Eq. \eqref{eq:5a} describes the cross-modal spatial alignment task. We optimize this by employing the gradient-based algorithm, and the $t$-th optimization stage can be expressed as:
\begin{equation}
\phi^{t} = \phi^{t-1} - \alpha \nabla_{\phi} \Psi(x^{t-1},\mathcal{T}(x_{\text{ref}}, \phi^{t-1})), \tag{6} \label{eq:6}
\end{equation}
where \( \alpha = \rho \lambda \), \(\rho\) is the stage size and \( \nabla_{\phi} \Psi(x^{t-1},\mathcal{T}(x_{\text{ref}}, \phi^{t-1})) \) is the gradient of the aligned cross-modal prior term with respect to the displacement field.

Eq. \eqref{eq:5b} represents the image reconstruction task. At this point, we regard the aligned cross-modal prior term as a regularization term. Given that the HQS algorithm has been proven effective for image inverse problems \citep{geman1992constrained,he2013half}, by introducing two auxiliary variables \( z \) and \( s \), we further transform Eq. \eqref{eq:5b} to an unconstrained optimization problem:
\begin{align}
\mathop{\min}_{x, s, z} 
&\frac{1}{2} \left\| F_m x-\tilde{k}\right\|^2+\lambda \Psi(z,\mathcal{T}(x_{\text{ref}}, \phi))+\eta R(s) \nonumber \\
&+ \frac{\beta_{1}}{2}\| x- z\|_2^2+ \frac{\beta_{2}}{2}\| x- s\|_2^2, \tag{7} \label{eq:7}
\end{align}
where \( \beta_1 \) and \( \beta_2 \) are penalty parameters. As \( \beta_1 \) and \( \beta_2 \) approach infinity, the result of minimizing Eq. \eqref{eq:7} will converge to the result of minimizing equation Eq. \eqref{eq:5b}. The optimization for the \( t \)-th iteration can be represented as:
\begin{align}
z^{t} &= \mathop{\arg\min}_{z} \frac{\beta_1}{2}\|{x}^{t-1} - {z}\|_2^2 + \lambda \Psi(z,\mathcal{T}(x_{\text{ref}}, \phi^{t})), \tag{8a} \label{eq:8a}\\
s^{t} &= \mathop{\arg\min}_{s} \frac{\beta_2}{2}\|{x}^{t-1} - {s}\|_2^2 + \eta R({s}), \tag{8b} \label{eq:8b}\\
x^{t} &= \mathop{\arg\min}_{x} \frac{1}{2} \left\| {F_m x} - \tilde{{k}} \right\|^2 + \frac{\beta_1}{2}\|{x} - {z}^{t}\|_2^2+\frac{\beta_2}{2}\|{x} - {s}^{t}\|_2^2. \tag{8c}\label{eq:8c}
\end{align}

\textbf{Update \(z\):} Given target modality image \(x^{t-1}\) and aligned reference modality image \(\mathcal{T}(x_{\text{ref}}, \phi^{t})\). We define the proximal operator \( \text{prox}_{\frac{\lambda}{\beta_1}\Psi(\cdot)}(\cdot) \) such that
\(\text{prox}_{\frac{\lambda}{\beta_1}\Psi(\cdot, \mathcal{T}(x_{\text{ref}}, \phi^{t}))}({x}) = \mathop{\arg\min}_{z} \frac{\beta_1}{2}\|  {x}-  {z}\|_2^2 + \lambda \Psi(  {z},\mathcal{T}(x_{\text{ref}}, \phi^{t}))\). Eq. \eqref{eq:8a} is then solved by the following equation:
\begin{equation}
z^{t} = \text{prox}_{\frac{\lambda}{\beta_1}\Psi(\cdot, \mathcal{T}(x_{\text{ref}}, \phi^{t}))}(x^{t-1}). \tag{9} \label{eq:9}
\end{equation}

\textbf{Update \(s\):} Given the target modality image \(x^{t-1}\). We also define the proximal operator \( \text{prox}_{\frac{\eta}{\beta_2}R}(\cdot) \) such that
\(
\text{prox}_{\frac{\eta}{\beta_2}R}({x}) = \mathop{\arg\min}_{s} \frac{\beta_2}{2}\|  {x}-  {s}\|_2^2+\eta R(  {s}).
\)
The solution to Eq. \eqref{eq:8b} can then be expressed as:
\begin{equation}
s^{t} = \text{prox}_{\frac{\eta}{\beta_2}R}(x^{t-1}). \tag{10} \label{eq:10}
\end{equation}

\textbf{Update \(x\):} Eq. \eqref{eq:8c} represents a quadratic regularized least squares problem and has a closed-form solution:
\begin{align}
x^{t}&=\left(F_m^H F_m+\left(\beta_1+\beta_2\right) I\right)^{-1} \left(F_m^H \tilde{k}+\beta_1 z^{t}+\beta_2 s^{t}\right) \nonumber\\ 
&=   {F}^H \Lambda^{-1}\left(M^H \tilde{k}+\beta_1 {F} z^{t} + \beta_2 {F} s^{t}\right),\tag{11} \label{eq:11}
\end{align}
where, \(\Lambda = \operatorname{diag}(\beta_1 + \beta_2)\), \( F_m^H \) is the Hermitian transpose of \( F_m \), and \( I \) is the identity matrix.

Through the iterative updates of \(\phi\), \(z\), \(s\), \(x\), we can address the joint alignment-reconstruction model. We unfold this optimization algorithm into a deep unfolding network and name it Deep Unfolding Network with Spatial Alignment (DUN-SA), as shown in Fig. \ref{fig1}. The details are
presented in Section \ref{DUNWSA}. 

\subsection{Deep Unfolding Network with Spatial Alignment}
\label{DUNWSA}
In Section \ref{SAAMCSA}, we propose an optimization algorithm for the joint alignment-reconstruction model. However, this algorithm has some limitations that affect its efficiency and performance. First, it requires integrating the spatial alignment task with the reconstruction process, which complicates the design of the aligned cross-modal prior term due to the need to effectively solve both tasks simultaneously. Second, similar to traditional optimization algorithms, this method requires dozens of iterations to converge, significantly increasing both the computational cost and the time to solution. To alleviate these problems, inspired by deep unfolding networks, we unfold the optimization algorithm into corresponding network modules.
\subsubsection{Model Overview}
\label{mo}
The core concept of the Deep Unfolding Network with Spatial Alignment is to solve the joint alignment-reconstruction model by unfolding the proposed algorithm with network modules and learning all parameters in an end-to-end manner.

The pipeline of the proposed DUN-SA is illustrated in Fig. \ref{fig1}. The entire network structure consists of several key modules. The Spatial Alignment Module (SAM) optimizes the displacement field, which explicitly contribute to the spatial alignment between different modalities. The Reconstruction Module (RM) aims to reconstruct the target modality image. Within RM, there is the Aligned Inter-modality Prior Learning Block (AIPLB), which is responsible for learning inter-modality prior. Additionally, the Denoising Block (DB) focuses on learning intra-modality denoising prior. Lastly, the Data Consistency Block (DCB) integrates both inter- and intra-modality prior to refine the reconstruction result. Notably, SAM and RM have a mutual improvement relationship. Specifically, SAM faces the challenge of aligning a fully-sampled reference image with an under-sampled target image, which may cause alignment errors due to artifacts. By utilizing intermediate reconstructions from RM, which progressively reduces such artifacts, SAM achieves more accurate alignments. Conversely, RM benefits from these improved alignments, which facilitate more precise inter-modality correspondences and result in higher quality reconstructions.
\begin{figure}[!ht]
\centerline{\includegraphics[width=9cm]{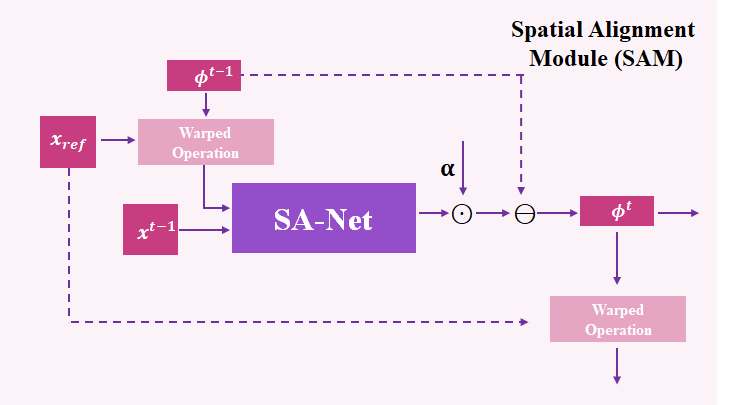}}
\caption{Architecture of Spatial Alignment Module (SAM).}
\label{fig2}
\end{figure}
\begin{figure*}[!ht]
\centerline{\includegraphics[width=\textwidth]{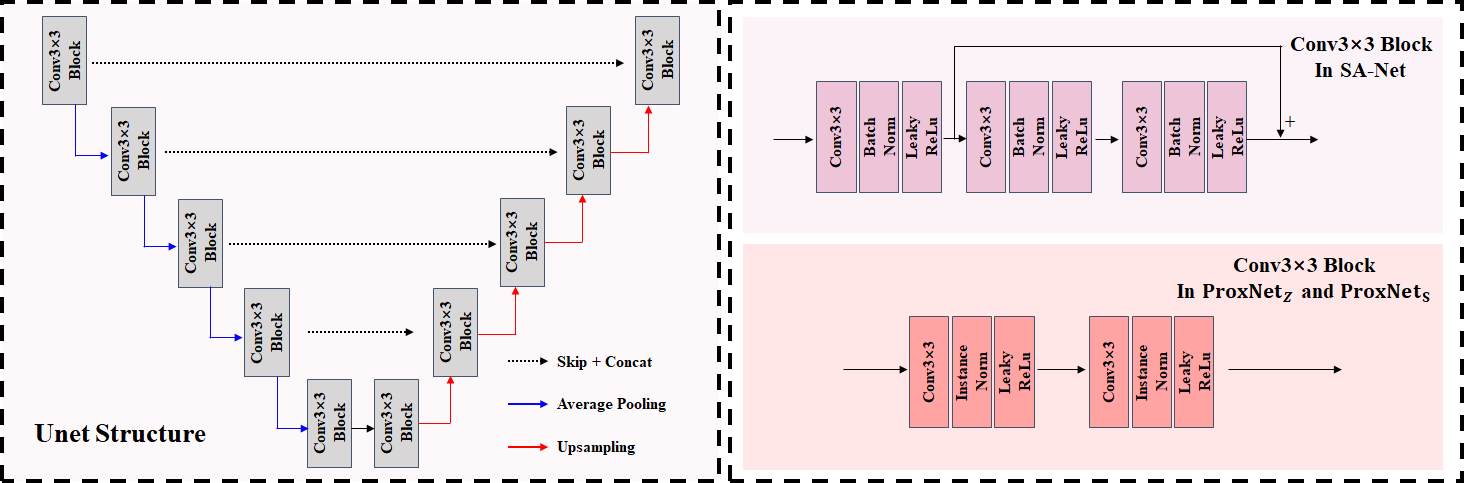}}
\caption{Detailed configurations of SA-Net, \texorpdfstring{$\text{ProxNet}_Z$}{\text{ProxNet}\_Z} and \texorpdfstring{$\text{ProxNet}_S$}{\text{ProxNet}\_S}.}
\label{fig3}
\end{figure*}

\textbf{SAM:} As shown in Fig. \ref{fig1}, the SAM refines the displacement field to improve spatial alignment between modalities. At stage \(t\), The network input is composed of the original reference modality image $x_{ref}$ (with the shape of $N_{x} \times N_{y}$), the displacement field from the previous stage $\phi^{t-1}$ (with the shape of $2 \times N_{x} \times N_{y}$), and the intermediate reconstructed image $x^{t-1}$ (with the shape of $N_{x} \times N_{y}$). The outputs are the refined displacement field $\phi^{t}$ (with the shape of $2 \times N_{x} \times N_{y}$) and the corresponding aligned reference modality image $\mathcal{T}(x_{\text{ref}},\phi^{t})$ (with the shape of $N_{x} \times N_{y}$). Specifically, SAM is built by unfolding the iterative rules in Eq. \eqref{eq:6}. It starts by warping the original reference modality image using $\phi^{t-1}$, and then concating with $x^{t-1}$ to execute the gradient operator $\nabla_{\phi} \Psi(\cdot)$. Subsequently, the displacement field is updated through gradient descent: \(\phi^{t} = \phi^{t-1} - \alpha \nabla_{\phi} \Psi(x^{t-1},\mathcal{T}(x_{\text{ref}}, \phi^{t-1}))\). Considering the complexity of the aligned cross-modal prior term, which makes calculating \(\nabla_{\phi} \Psi(\cdot)\) difficult, we use SA-Net (Spatial Alignment Network) to replace the gradient operator, resulting in the update \(\phi^{t}=\phi^{t-1}- \alpha \text{SA-Net}(  {x^{t-1}},\mathcal{T}(x_{\text{ref}}, \phi^{t-1}))\). Notably, we initialize $\phi^{0}$ with identity transformation based on the shape of $x_{ref}$. The architecture of SAM is shown in Fig. \ref{fig2}. Details on SA-Net are further discussed in Section \ref{SA2M2LB}.

\textbf{RM:} At stage \(t\), the inputs to the Reconstruction Module (RM) include the intermediate reconstructed image \(x^{t-1}\) (with the shape of $N_{x} \times N_{y}$) and the aligned reference modality image \(\mathcal{T}(x_{\text{ref}}, \phi^{t-1})\) (with the shape of $N_{x} \times N_{y}$) provided by SAM. The RM aims to enhance the clarity and quality of the image, producing an improved intermediate reconstruction image \(x^{t}\) (with the shape of $N_{x} \times N_{y}$) by solving Eq. \eqref{eq:5b}. Specifically, we utilize HQS algorithm as detailed in Section \ref{toa}, to effectively tackle this equation, and design AIPLB, DB and DCB correspondingly.

\textbf{AIPLB, DB, and DCB:} The framework utilizes AIPLB, DB, and DCB to learn inter-modality prior, intra-modality prior, and to refine the reconstruction of the target modality image, respectively. As shown in Fig. \ref{fig1}, at stage \(t\), \(x^{t-1}\) is concatenated with \(\mathcal{T}(x_{\text{ref}},\phi^{t})\) to serve as input into AIPLB for learning inter-modality prior \(z^{t}\) (with the shape of $N_{x} \times N_{y}$). Meanwhile, \(x^{t-1}\) is fed into DB to learn intra-modality prior \(s^{t}\) (with the shape of $N_{x} \times N_{y}$). Subsequently, both types of priors are input into DCB along with under-sampled target modality k-space data \(\mathop{\tilde{k}}\) for further refinement. Specifically, AIPLB is constructed by unfolding Eq. \eqref{eq:9} and using $\text{ProxNet}_Z$ to replace the proximal operator \( \text{prox}_{\frac{\lambda}{\beta_1} \Psi(\cdot)}(\cdot) \). Furthermore, to stabilize network training, we enable $\text{ProxNet}_Z$ to predict the residual by adding a skip connection from the input to the output, thereby obtaining \( z^{t}=x^{t-1}+\text{$\text{ProxNet}_Z$}(x^{t-1},\mathcal{T}(x_{\text{ref}}, \phi^{t})) \). Similarly, for learning the intra-modality prior in DB, we unfold Eq. \eqref{eq:10} and use $\text{ProxNet}_S$ to replace the proximal operator \( \text{prox}_{\frac{\eta}{\beta_2}R}(\cdot) \), we also enable $\text{ProxNet}_S$ to predict the residual by adding a skip connection from the input to the output. Finally, in DCB, we unfold Eq. \eqref{eq:11} and use the obtained inter- and intra-modality prior \(z^{t}\) and \(s^{t}\), together with under-sampled target modality k-space data \(\mathop{\tilde{k}}\), to refine the reconstruction results, obtaining \(x^{t} = {F}^H \Lambda^{-1}\left(M^H \mathop{\tilde{k}} + \beta_1 {F} z^{t} + \beta_2 {F} s^{t}\right).\) With the k-stage optimization, the proposed DUN-SA can finely reconstruct the target modality image. Further details about network implementation are described in Section \ref{SA2M2LB}.

In this way, the proposed DUN-SA is correspondingly constructed under the guidance of the optimization algorithm, and every network module has its own physical meaning, corresponding to specific iterative step. It is worth noting that we do not need to explicitly calculate the gradient of the aligned cross-modal prior term and regularization term. Instead, the network learns the updates of $\phi$, $z$, $s$ at each stage. Therefore, we can optimize the process of the optimization algorithm by updating the network parameters.

\subsubsection{SA-Net, \texorpdfstring{$\text{ProxNet}_Z$ and $\text{ProxNet}_S$}{\text{ProxNet}\_Z and \text{ProxNet}\_S}}
\label{SA2M2LB}
In SAM, the displacement field \(\phi^{t}\) is updated as follows: \(\phi^{t}=\phi^{t-1}- \alpha \text{SA-Net}(  {x^{t-1}},\mathcal{T}(x_{\text{ref}}, \phi^{t-1}))\). This update aims to estimate and compensate for the subtle spatial misalignments between the target and reference modalities. To effectively capture the multi-scale information between images, we employ a CNN architecture known as U-Net \citep{10.1007/978-3-319-24574-4_28} as the backbone of SA-Net. The learnable parameters include the scaling factor $\alpha$, which is initialized to be 1, and the parameters within the SA-Net, which is initialized to be 0. The SA-Net consists of an encoder and a decoder. As shown in Fig. \ref{fig3}, the encoder consists of four encoding blocks with residual connection, which contain three convolutional layers with 3 × 3 kernels, Batch Normalization layers and Leaky ReLU nonlinearity. Corresponding to the encoder, the decoder also consists of four decoding blocks with residual connection, which contain three convolutional layers with 3 × 3 kernels, Batch Normalization layers and Leaky ReLU nonlinearity. Average pooling and nearest-neighbor upsampling are adopted to change the spatial size of feature maps. 

In AIPLB and DB, instead of utilizing hand-crafted regularization to learn inter-modality and intra-modality prior, we substitute the proximal operator with learnable networks, namely $\text{ProxNet}_Z$ and $\text{ProxNet}_S$ as detailed in Section \ref{mo}. 

For $\text{ProxNet}_Z$, previous studies \citep{10.1007/978-3-030-00928-1_25,8758456,8552399,9115255} have underscored the effectiveness of the U-Net architecture in extracting features from multi-modal data. This process involves concatenating aligned images from two modalities along the channel dimension and inputting them into the network, where the forward convolutional layers amalgamate information and features from corresponding channel positions. In this paper, we utilize a variant of U-Net as the backbone architecture for $\text{ProxNet}_Z$.

For $\text{ProxNet}_S$, we consider Eq. \eqref{eq:10} as a denoising process driven by the intra-modality denoising prior \citep{8434321,pmlr-v172-xin22a,8550778}. Typically, any established image denoising network could be applied here. Given the simplicity and proven effectiveness of U-Net in learning intra-modality denoising prior \citep{10.1007/978-3-030-59713-9_7,pmlr-v172-xin22a}, we adopted the same structural framework for $\text{ProxNet}_S$ as used for $\text{ProxNet}_Z$, but with different inputs, as pointed out in Section \ref{mo}.

Both $\text{ProxNet}_Z$ and $\text{ProxNet}_S$ consist of an encoder and a decoder. As shown in Fig. \ref{fig3}, the encoder consists of four encoding blocks, which contain two convolutional layers with 3 × 3 kernels, Instance Normalization layers and Leaky ReLU nonlinearity. Corresponding to the encoder, the decoder also consists of four decoding blocks with residual connection, which contain two convolutional layers with 3 × 3 kernels, Instance Normalization layers and Leaky ReLU nonlinearity. Average pooling and transpose convolutional upsampling are adopted to change the spatial size of feature maps.

\subsection{Network parameters and network training}
\label{NPLF}
Network parameters are categorized into two sets; $\alpha$ and SA-Net parameters in SAM are denoted as $\omega$, which are utilized to solve the spatial alignment task. Parameters for $\text{ProxNet}_Z$ in AIPLB, $\text{ProxNet}_S$ in the DB, and $\beta_1$, $\beta_2$ in DCB, which address the reconstruction task, are collectively denoted as $\theta$. Inspired by \cite{9745968}, in our network training, we use the Structural Similarity Index Measure (SSIM) as our final loss function:
\[
\mathcal{L} (\omega, \theta) = \frac{1}{N} \sum_{i=1}^N (1 - \text{SSIM}(x_{\text{rec},i} (\omega, \theta), x_{\text{gt},i})) \tag{17}
\]
where $x_{\text{rec},i}$ denotes the $i^{th}$ image reconstructed by the network, and $x_{\text{gt},i}$ signifies the $i^{th}$ ground-truth image. 

For $\omega$: we fix $\theta$ and update $\omega$ as follows:
\[
\omega \leftarrow \omega - \eta_1 \nabla_{\omega} \mathcal{L} (\omega, \theta^*)
\]
where \(\eta_1\) is the learning rate, and \(\nabla_{\omega} \mathcal{L}\) represents the gradient of the loss function with respect to \(\omega\).

For $\theta$: While keeping $\omega$ fixed, $\theta$ is updated as follows:
\[
\theta \leftarrow \theta - \eta_2 \nabla_{\theta} \mathcal{L} (\omega^*, \theta)
\]
where \(\eta_2\) is the learning rate, and \(\nabla_{\omega} \mathcal{L}\) represents the gradient of the loss function with respect to \(\theta\).

\section{Experiments settings}
\subsection{Datasets}
We evaluate our method using four datasets, namely the fastMRI dataset, the IXI dataset, an In-house dataset and the BraTS 2018 dataset.

\textbf{FastMRI Dataset\footnote{\url{https://fastMRI.med.nyu.edu/}.}}: To maintain consistency in the experiments, we follow the setups described in \citep{9745968} and select 340 pairs of T1-weighted and T2-weighted axial brain MRIs. We use 170 volumes (2720 pairs of slices) as the training set, 68 volumes (1088 pairs of slices) as the validation set, and 102 volumes (1632 pairs of slices) as the test set. The in-plane size of all T1-weighted images and T2-weighted images is \(320 \times 320\), with the resolution of \(0.68mm \times 0.68mm\) and the slice spacing of 5mm.
    
\textbf{IXI Dataset\footnote{\url{http://brain-development.org/ixi-dataset/}.}}: The IXI dataset contains 576 paired multi-modal 3D brain MRIs. We select 570 pairs of PD-weighted and T2-weighted axial brain MRIs for our experiments. Specifically, we use 285 volumes as the training set, 115 volumes as the validation set, and 170 volumes as the test set. The in-plane size of all PD-weighted images and T2-weighted images is \(256 \times 256\). Consistent with \citep{ijcai2023p112}, we select the middle 100 slices from each volume for our experiments.

\textbf{In-house Dataset}: The In-house dataset contains 3D brain MRIs of 34 subjects, including paired T1-weighted images and T2-weighted images of their whole brains. Among them, 24 pairs of volumes (1797 pairs of slices) are employed as the training set. 4 pairs of volumes (300 pairs of slices) refer to the validation set, and 6 pairs of volumes (450 pairs of slices) are used for testing. The in-plane size of all T1-weighted images and T2-weighted images is \(320 \times 320\), with the resolution of \(0.68mm \times 0.68mm\) and the slice spacing of 5mm.

\textbf{BraTS 2018 Dataset\footnote{\url{https://ipp.cbica.upenn.edu/}.}}: The BraTS 2018 dataset includes clinically-acquired, multi-modal MRI scans from patients with glioblastoma, featuring four different imaging modalities: T1, T2, Flair, and T1-weighted contrast-enhanced (T1CE). We selected the high-grade gliomas (HGG) subset, which comprises 210 3D brain MRIs. Specifically, we use 105 volumes as the training set, 42 volumes as the validation set, and 63 volumes as the test set. The in-plane size of all images is \(240 \times 240\), and we select the middle 20 slices from each volume for our experiments.

\subsection{Compared Methods}
\begin{table*}[!ht]
    \centering
    \small
    \caption{Quantitative evaluation of DUN-SA vs. other methods on the fastMRI dataset for 4$\times$ and 8$\times$ acceleration under \textbf{equispaced} and \textbf{random} 1D subsampling masks, where T1-weighted images are used as reference modality to assist the reconstruction of T2-weighted images. Best results are emphasized in \textbf{bold}, and the second best are emphasized with an \underline{underline}.}
    \renewcommand{\arraystretch}{0.93}
    \begin{tabular}{l *{7}{c}}
        \toprule
        &\multirow{2}{*}{\textbf{Methods}} & \multicolumn{3}{c}{\textbf{ 4$\times$ Acceleration}} & \multicolumn{3}{c}{\textbf{8$\times$ Acceleration}} \\
        \cmidrule(lr){3-5} \cmidrule(lr){6-8} &
        & \textbf{PSNR} & \textbf{SSIM} & \textbf{MAE} & \textbf{PSNR} & \textbf{SSIM} & \textbf{MAE} \\
        \midrule
        \multirow{8}{*}{\rotatebox[origin=c]{90}{\textbf{Equispaced}}} &Zero-filling & 26.92$\pm$1.02 & 0.7321$\pm$0.0279 & 0.0238$\pm$0.0034 & 24.36$\pm$1.11 & 0.6428$\pm$0.0350 & 0.0328$\pm$0.0052 \\
        &E2E-Varnet & 38.81$\pm$1.61 & 0.9761$\pm$0.0065 & 0.0058$\pm$0.0012 & 36.66$\pm$1.65 & 0.9658$\pm$0.0087 & 0.0074$\pm$0.0015 \\
        &HQS-Unet & 39.31$\pm$1.69 & 0.9773$\pm$0.0063 & 0.0056$\pm$0.0011 & 37.12$\pm$1.68 & 0.9683$\pm$0.0085 & 0.0071$\pm$0.0017 \\
        &MD-DUN & 40.31$\pm$1.84 & 0.9721$\pm$0.0069 & 0.0053$\pm$0.0012 & 38.45$\pm$1.78 & 0.9618$\pm$0.0096 & 0.0061$\pm$0.0014 \\
        &MM-E2E-Varnet & 40.28$\pm$1.80 & 0.9806$\pm$0.0067 & 0.0051$\pm$0.0011 & 38.35$\pm$1.89 & 0.9728$\pm$0.0091 & 0.0063$\pm$0.0015 \\
        &SAN & 40.48$\pm$1.77 & \underline{0.9819$\pm$0.0065} & 0.0048$\pm$0.0011 & \underline{38.93$\pm$1.87} & \underline{0.9764$\pm$0.0082} & \underline{0.0059$\pm$0.0014} \\
        &MC-CDic & \underline{40.72$\pm$1.80} & 0.9814$\pm$0.0067 & \underline{0.0047$\pm$0.0011} & 38.25$\pm$1.85 & 0.9724$\pm$0.0100 & 0.0062$\pm$0.0014 \\
        &DUN-SA & \textbf{41.48$\pm$1.84} & \textbf{0.9838$\pm$0.0060} & \textbf{0.0045$\pm$0.0010} & \textbf{40.23$\pm$1.88} & \textbf{0.9802$\pm$0.0072} & \textbf{0.0052$\pm$0.0012} \\
        \midrule
        \multirow{8}{*}{\rotatebox[origin=c]{90}{\textbf{Random}}}&Zero-filling & 27.12$\pm$1.03 & 0.7373$\pm$0.0282 & 0.0233$\pm$0.0032 & 23.88$\pm$1.70 & 0.5818$\pm$0.0431 & 0.0342$\pm$0.0067 \\
        &E2E-Varnet & 42.60$\pm$1.71 & 0.9868$\pm$0.0044 & 0.0041$\pm$0.0009 & 35.82$\pm$1.63 & 0.9612$\pm$0.0094 & 0.0080$\pm$0.0017 \\
        &HQS-Unet & 43.05$\pm$1.87 & 0.9873$\pm$0.0042 & 0.0040$\pm$0.0008 & 36.22$\pm$1.73& 0.9637$\pm$0.0093 & 0.0077$\pm$0.0016 \\
        &MD-DUN & 43.45$\pm$1.80 & 0.9832$\pm$0.0068 & 0.0037$\pm$0.0008 & 37.28$\pm$1.86 & 0.9541$\pm$0.0099 & 0.0069$\pm$0.0017 \\
        &MM-E2E-Varnet & 43.47$\pm$1.82 & 0.9879$\pm$0.0045 & 0.0038$\pm$0.0009 & 37.40$\pm$1.89 & 0.9705$\pm$0.0093 & 0.0071$\pm$0.0015 \\
        &SAN & 43.87$\pm$1.82 & \underline{0.9886$\pm$0.0043} & 0.0038$\pm$0.0009 & \underline{37.84$\pm$1.94} & \underline{0.9724$\pm$0.0091} & 0.0069$\pm$0.0015 \\
        &MC-CDic & \underline{44.01$\pm$1.82} & 0.9885$\pm$0.0038 & \underline{0.0036$\pm$0.0008} & 37.60$\pm$1.82 & 0.9705$\pm$0.0100 & \underline{0.0067$\pm$0.0015} \\
        &DUN-SA & \textbf{45.06$\pm$1.85} & \textbf{0.9907$\pm$0.0029} & \textbf{0.0032$\pm$0.0007} & \textbf{39.22$\pm$1.84} & \textbf{0.9770$\pm$0.0075} & \textbf{0.0057$\pm$0.0012} \\
        \bottomrule
    \end{tabular}
    \label{tb1}
\end{table*}
To evaluate the performance of our model (DUN-SA), we conduct comparisons with state of the art methods, including two single-modal MRI reconstruction methods: E2E-Varnet \citep{10.1007/978-3-030-59713-9_7}, HQS-Unet \citep{pmlr-v172-xin22a}; and four multi-modal MRI reconstruction methods: MD-DUN \citep{10.1007/978-3-030-59713-9_19}, MM-E2E-Varnet(a variant of E2E-VarNet for multi-modal input), SAN \citep{9745968}, and MC-CDic \citep{ijcai2023p112}. The details are presented as follows.

\textbf{E2E-Varnet} employs the variational technique to solve the MRI reconstruction problem, unfolding it into a network structure combined with U-Net for end-to-end learning. 

\textbf{HQS-Unet} utilizes the HQS algorithm, integrating U-Net with residual and buffering designs to learn a denoising prior in an end-to-end manner. 

\textbf{MD-DUN} builds upon the unfolded HQS algorithm and incorporates spatial and channel attention mechanisms within the denoising module, enabling more effective integration of reference modality information. 

\textbf{MM-E2E-Varnet} modifies the original E2E-Varnet by replacing its architecture with a variant of U-Net consistent with multi-modal input, and utilizes reference modality information. 

\textbf{SAN} performs image alignment between the target and reference modalities in advance, more effectively utilizing the information of the reference modality.

\textbf{MC-CDic} leverages the convolutional dictionary learning-based model and proximal gradient algorithm to propose a corresponding multi-scale convolutional dictionary network for multi-modal MRI reconstruction.

\subsection{Performance Evaluation Metric}
To accurately and comprehensively evaluate the performance of image reconstruction, we select three core metrics: Peak Signal-to-Noise Ratio (PSNR), Structural Similarity Index Measure (SSIM), and Mean Absolute Error (MAE). These metrics are employed to quantify the similarity between the reconstructed images and the corresponding ground truth. It is noteworthy that higher values of PSNR and SSIM, along with lower MAE values, indicate superior reconstruction performance of the model.

\subsection{Implementation Details}
In our experiments, we follow the paradigm set by the fastMRI challenge, where under-sampled MRIs are obtained by masking the corresponding fully-sampled MRIs in k-space using a Cartesian sampling pattern. Specifically, we employ two types of sampling patterns: random and equispaced, to thoroughly assess and compare the performance of DUN-SA with other methods, given their distinct impacts on reconstruction. These are applied at sampling ratios of 25\% (4$\times$ acceleration) and 12.5\% (8$\times$ acceleration), respectively. Acknowledging that low-frequency signals contain the majority of energy in k-space, we allocate 32\% of the sampling to these low frequencies in both patterns. The remaining portion of the sampling is distributed either randomly or in an equispaced manner. 

Concerning the selection of hyperparameters for the network, we choose a model size with stage$=$12 for our experiments.

All network models used in these experiments are implemented using the PyTorch framework and are run on a computer equipped with four NVIDIA GeForce GTX 3090 GPUs. During the model training phase, we set the batch size to be \(2\), adopt an end-to-end training strategy, and use the SSIM loss function with a coefficient of \(1\). We train and test using different slice sizes for each dataset: (320, 320) for the fastMRI and In-house datasets, (256, 256) for the IXI dataset, and (240, 240) for the BraTs 2018 dataset. All parameters are optimized using the Adam optimizer with a learning rate of \(1 \times 10^{-4}\). To prevent overfitting, we used a validation set to select the optimal parameters during the training process. The final experimental results are based on these optimal parameters, as selected and validated through the validation set, and applied on the test set.

\begin{figure*}[!ht!ht]
\centerline{\includegraphics[width=\textwidth]{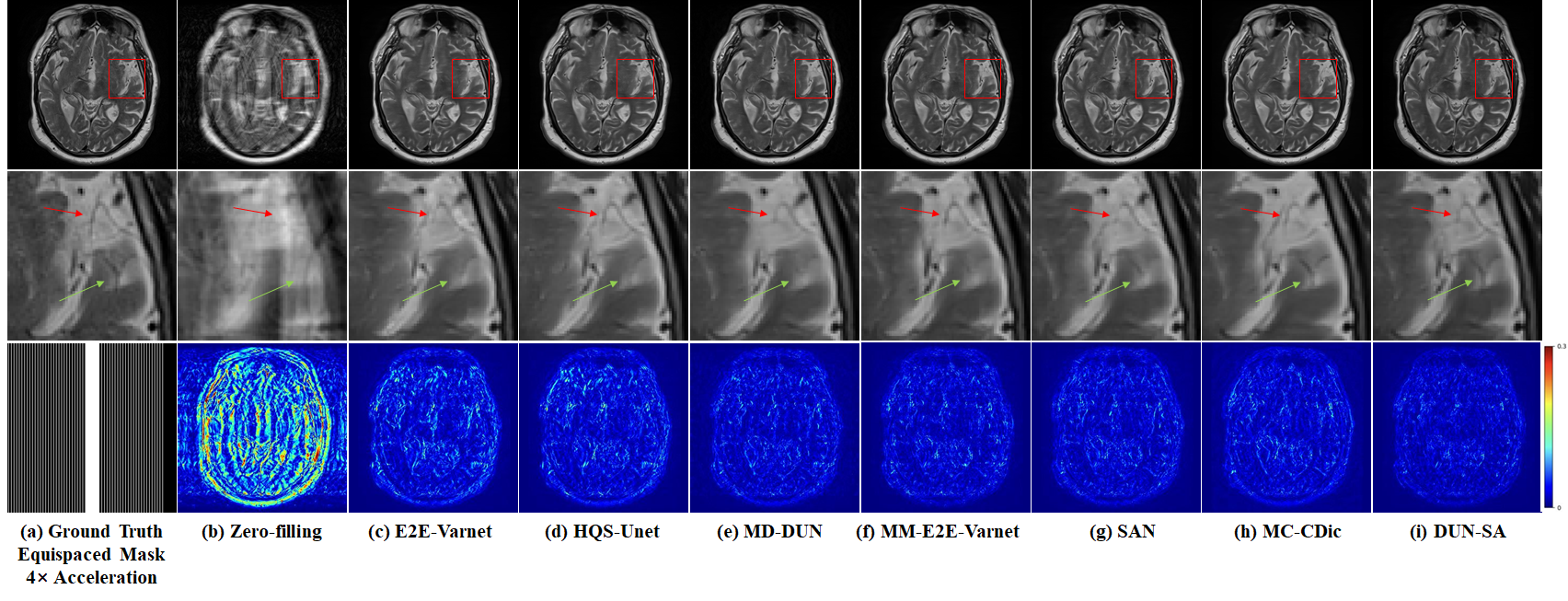}}
\caption{Visual comparison with representative methods for 4$\times$ acceleration under 1D equispaced subsampling mask on fastMRI dataset. First row: Reconstructed images by different methods; second row: Zoomed-in region of interest; third row: Equispaced mask of 4$\times$ acceleration and error maps of different methods.}
\label{fig4}
\end{figure*}

\section{Experimental results}
\subsection{Results on the fastMRI dataset}
\label{efd}

\begin{table*}[!ht]
    \centering
    \small
    \caption{Quantitative evaluation of DUN-SA vs. other methods on the IXI dataset for 4$\times$ and 8$\times$ acceleration under \textbf{equispaced} and \textbf{random} 1D subsampling masks, where PD-weighted images are used as reference modality to assist the reconstruction of T2-weighted images. Best results are emphasized in \textbf{bold}, and the second best are emphasized with an \underline{underline}.}
    \renewcommand{\arraystretch}{0.93}
    \begin{tabular}{l *{7}{c}}
        \toprule
        &\multirow{2}{*}{\textbf{Methods}} & \multicolumn{3}{c}{\textbf{ 4$\times$ Acceleration}} & \multicolumn{3}{c}{\textbf{8$\times$ Acceleration}} \\
        \cmidrule(lr){3-5} \cmidrule(lr){6-8} &
        & \textbf{PSNR} & \textbf{SSIM} & \textbf{MAE} & \textbf{PSNR} & \textbf{SSIM} & \textbf{MAE} \\
        \midrule
        \multirow{8}{*}{\rotatebox[origin=c]{90}{\textbf{Equispaced}}} &Zero-filling & 26.56$\pm$2.04 & 0.6337$\pm$0.0449 & 0.0287$\pm$0.0064 & 24.08$\pm$2.01 & 0.5517$\pm$0.0535 & 0.0369$\pm$0.0084 \\
        &E2E-Varnet & 41.87$\pm$2.29 & 0.9859$\pm$0.0044 & 0.0042$\pm$0.0011 & 34.51$\pm$2.10 & 0.9521$\pm$0.0109 & 0.0095$\pm$0.0022 \\
        &HQS-Unet & 42.73$\pm$2.33 & 0.9874$\pm$0.0036 & 0.0040$\pm$0.0009 & 35.41$\pm$2.12 & 0.9575$\pm$0.0090 & 0.0086$\pm$0.0020 \\
        &MD-DUN & 45.17$\pm$2.39 & 0.9874$\pm$0.0038 & 0.0034$\pm$0.0009 & 40.52$\pm$2.21 & 0.9729$\pm$0.0065 & 0.0058$\pm$0.0014 \\
        &MM-E2E-Varnet & 45.96$\pm$2.40 & 0.9921$\pm$0.0030 & 0.0030$\pm$0.0008 & 41.19$\pm$2.25 & 0.9832$\pm$0.0056 & 0.0049$\pm$0.0012 \\
        &SAN & \underline{46.17$\pm$2.44} & \underline{0.9923$\pm$0.0030} & \underline{0.0029$\pm$0.0007} & \underline{41.21$\pm$2.27} & \underline{0.9833$\pm$0.0057} & 0.0049$\pm$0.0012 \\
        &MC-CDic & 45.79$\pm$2.41 & 0.9916$\pm$0.0032 & 0.0031$\pm$0.0006 & 40.82$\pm$2.23 & 0.9823$\pm$0.0060 & \underline{0.0048$\pm$0.0012} \\
        &DUN-SA & \textbf{47.25$\pm$2.53} & \textbf{0.9936$\pm$0.0026} & \textbf{0.0025$\pm$0.0007} & \textbf{41.84$\pm$2.33} & \textbf{0.9850$\pm$0.0053} & \textbf{0.0046$\pm$0.0011} \\
        \midrule
        \multirow{8}{*}{\rotatebox[origin=c]{90}{\textbf{Random}}}&Zero-filling & 26.32$\pm$2.04 & 0.6177$\pm$0.0465 & 0.0296$\pm$0.0068 & 23.91$\pm$2.01 & 0.5371$\pm$0.0538 & 0.0380$\pm$0.0086 \\
        &E2E-Varnet & 40.10$\pm$2.20 & 0.9808$\pm$0.0055 & 0.0054$\pm$0.0013 & 33.33$\pm$2.10 & 0.9417$\pm$0.0126 & 0.0107$\pm$0.0025 \\
        &HQS-Unet & 41.06$\pm$2.23 & 0.9838$\pm$0.0044 & 0.0049$\pm$0.0011 & 34.23$\pm$2.05 & 0.9486$\pm$0.0116 & 0.0098$\pm$0.0022 \\
        &MD-DUN & 43.82$\pm$2.30 & 0.9844$\pm$0.0053 & 0.0038$\pm$0.0010 & 39.81$\pm$2.18 & 0.9699$\pm$0.0067 & 0.0060$\pm$0.0015 \\
        &MM-E2E-Varnet & \underline{44.77$\pm$2.41} & 0.9905$\pm$0.0034 & 0.0034$\pm$0.0009 & 40.89$\pm$2.16 & 0.9821$\pm$0.0061 & 0.0052$\pm$0.0012 \\
        &SAN & 44.75$\pm$2.46 & \underline{0.9907$\pm$0.0034} & \underline{0.0034$\pm$0.0009} & \underline{41.09$\pm$2.26} & \underline{0.9829$\pm$0.0058} & \underline{0.0050$\pm$0.0012} \\
        &MC-CDic & 44.39$\pm$2.36 & 0.9877$\pm$0.0037 & 0.0037$\pm$0.0010 & 40.78$\pm$2.23 & 0.9792$\pm$0.0062 & 0.0052$\pm$0.0012 \\
        &DUN-SA & \textbf{45.38$\pm$2.51} & \textbf{0.9915$\pm$0.0032} & \textbf{0.0032$\pm$0.0008} & \textbf{41.62$\pm$2.29} & \textbf{0.9846$\pm$0.0054} & \textbf{0.0047$\pm$0.0012} \\
        \bottomrule
    \end{tabular}
    \label{tb2}
\end{table*}

\begin{figure*}[!ht]
\centerline{\includegraphics[width=\textwidth]{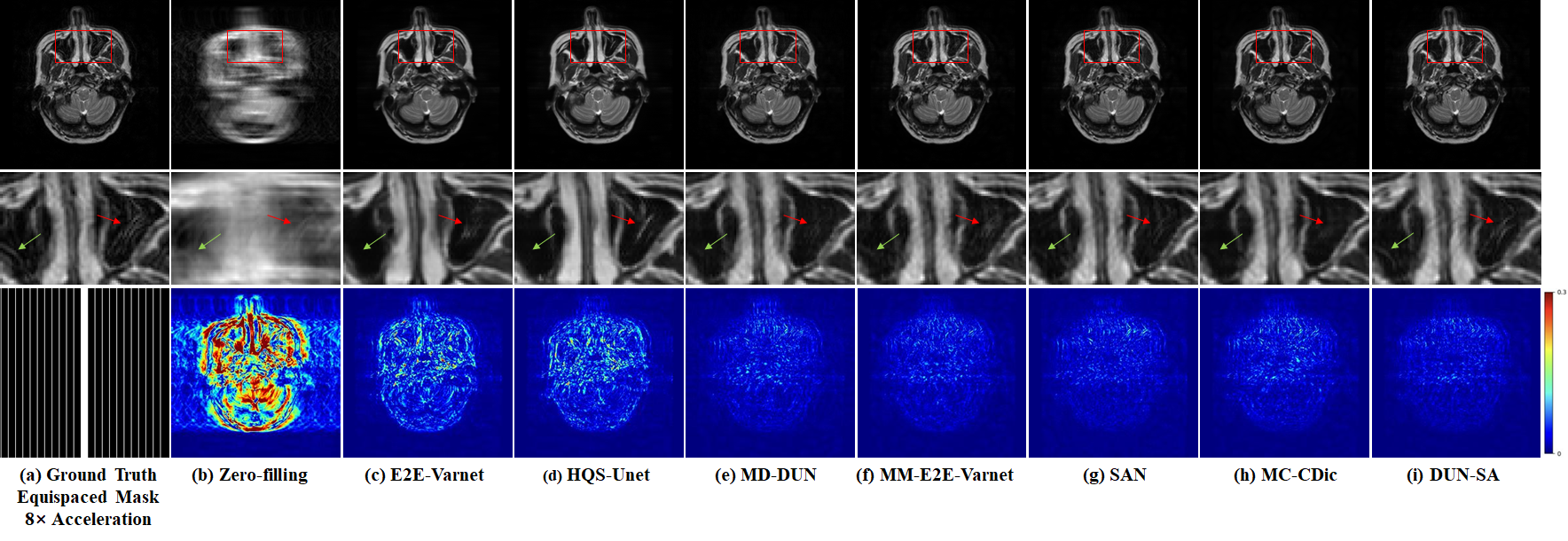}}
\caption{Visual comparison with representative methods for 8$\times$ acceleration under 1D equispaced subsampling mask on the IXI dataset. First row: Reconstructed images by different methods; second row: Zoomed-in region of interest; third row: Equispaced mask of 8$\times$ acceleration and error maps of different methods.}
\label{fig5}
\end{figure*}
We first run our proposed method on the fastMRI dataset. In this part of the experiments, we use the fully-sampled T1-weighted image as the reference modality to assist the k-space under-sampled data of the T2 modality in reconstruction.

Table \ref{tb1} displays the quantitative evaluations on the fastMRI dataset for 1D equispaced and random masks under 4× and 8× acceleration. We observe that images reconstructed by multi-modal methods show significant improvements in the PSNR, SSIM, and MAE metrics across all experimental setups compared to those reconstructed by single-modal methods. This enhancement is particularly pronounced for 8$\times$ acceleration. The introduction of spatial alignment in the multi-modal reconstruction further improves the PSNR, SSIM, and MAE values. It is evident that DUN-SA outperforms all others in every metric across all settings. In comparison with the second-best values, for 8$\times$ acceleration under random 1D subsampling masks, the highest differences in PSNR, SSIM, and MAE are 1.38db, 0.0046, and 0.001, respectively. In contrast, for 4$\times$ acceleration under equispaced 1D subsampling masks, the smallest differences are 0.76db, 0.0019, and 0.0002. These results suggest that DUN-SA can effectively utilize the reference modality to aid the MRI reconstruction of the target modality, leading to a noticeable improvement in reconstruction quality.

\begin{table*}[!ht]
    \centering
    \small
    \caption{Quantitative evaluation of DUN-SA vs. other methods on the In-house dataset for 4$\times$ and 8$\times$ acceleration under \textbf{equispaced} and \textbf{random} 1D subsampling masks, where T1-weighted images are used as reference modality to assist the reconstruction of T2-weighted images. Best results are emphasized in \textbf{bold}, and the second best are emphasized with an \underline{underline}.}
    \renewcommand{\arraystretch}{0.93}
    \begin{tabular}{l *{7}{c}}
        \toprule
        &\multirow{2}{*}{\textbf{Methods}} & \multicolumn{3}{c}{\textbf{ 4$\times$ Acceleration}} & \multicolumn{3}{c}{\textbf{8$\times$ Acceleration}} \\
        \cmidrule(lr){3-5} \cmidrule(lr){6-8} &
        & \textbf{PSNR} & \textbf{SSIM} & \textbf{MAE} & \textbf{PSNR} & \textbf{SSIM} & \textbf{MAE} \\
        \midrule
        \multirow{8}{*}{\rotatebox[origin=c]{90}{\textbf{Equispaced}}} &Zero-filling & 26.63$\pm$0.67 & 0.6864$\pm$0.0153 & 0.0272$\pm$0.0026 & 24.32$\pm$0.66 & 0.5962$\pm$0.0152 & 0.0362$\pm$0.0033 \\
        &E2E-Varnet & 35.42$\pm$0.77 & 0.9643$\pm$0.0031 & 0.0083$\pm$0.0009 & 33.90$\pm$0.79 & 0.9526$\pm$0.0044 & 0.0100$\pm$0.0011 \\
        &HQS-Unet & 35.62$\pm$0.79 & 0.9651$\pm$0.0032 & 0.0082$\pm$0.0009 & 34.13$\pm$0.77 & 0.9535$\pm$0.0043 & 0.0098$\pm$0.0011 \\
        &MD-DUN & 36.19$\pm$1.24 & 0.9587$\pm$0.0059 & 0.0079$\pm$0.0011 & 34.68$\pm$1.33 & 0.9451$\pm$0.0090 & 0.0092$\pm$0.0017 \\
        &MM-E2E-Varnet & 36.17$\pm$1.24 & 0.9684$\pm$0.0062 & 0.0078$\pm$0.0012 & 34.88$\pm$1.38 & 0.9600$\pm$0.0092 & 0.0090$\pm$0.0017 \\
        &SAN & \underline{36.58$\pm$1.25} & \underline{0.9705$\pm$0.0055} & \underline{0.0075$\pm$0.0012} & \underline{35.17$\pm$1.71} & \underline{0.9613$\pm$0.0112} & \underline{0.0089$\pm$0.0019} \\
        &MC-CDic & 36.21$\pm$1.28 & 0.9682$\pm$0.0064 & 0.0078$\pm$0.0013 & 34.55$\pm$1.36 & 0.9574$\pm$0.0097 & 0.0094$\pm$0.0018 \\
        &DUN-SA & \textbf{37.41$\pm$1.08} & \textbf{0.9747$\pm$0.0039} & \textbf{0.0069$\pm$0.0009} & \textbf{36.31$\pm$1.41} & \textbf{0.9687$\pm$0.0070} & \textbf{0.0078$\pm$0.0014} \\
        \midrule
        \multirow{8}{*}{\rotatebox[origin=c]{90}{\textbf{Random}}}&Zero-filling & 26.82$\pm$0.66 & 0.7075$\pm$0.0144 & 0.0261$\pm$0.0024 & 24.27$\pm$0.66 & 0.5982$\pm$0.0164 & 0.0359$\pm$0.0033 \\
        &E2E-Varnet & 37.57$\pm$0.78 & 0.9743$\pm$0.0020 & 0.0068$\pm$0.0007 & 32.31$\pm$0.78 & 0.9441$\pm$0.0044 & 0.0124$\pm$0.0013 \\
        &HQS-Unet & 37.88$\pm$0.77 & 0.9757$\pm$0.0020 & 0.0067$\pm$0.0007 & 32.64$\pm$0.76 & 0.9457$\pm$0.0043 & 0.0120$\pm$0.0012 \\
        &MD-DUN & 38.44$\pm$1.19 & 0.9702$\pm$0.0035 & 0.0065$\pm$0.0009 & 33.41$\pm$1.43 & 0.9386$\pm$0.0123 & 0.107$\pm$0.0021 \\
        &MM-E2E-Varnet & 38.61$\pm$1.18 & 0.9789$\pm$0.0038 & 0.0064$\pm$0.0010 & 33.46$\pm$1.41 & 0.9491$\pm$0.0123 & 0.0107$\pm$0.0020 \\
        &SAN & \underline{38.92$\pm$1.23} & \underline{0.9799$\pm$0.0038} & \underline{0.0061$\pm$0.0010} & \underline{33.74$\pm$1.53} & \underline{0.9511$\pm$0.0136} & \underline{0.0102$\pm$0.0023} \\
        &MC-CDic & 38.84$\pm$1.20 & 0.9797$\pm$0.0037 & 0.0062$\pm$0.0009 & 33.20$\pm$1.43 & 0.9484$\pm$0.0128 & 0.0110$\pm$0.0022 \\
        &DUN-SA & \textbf{40.47$\pm$1.09} & \textbf{0.9845$\pm$0.0024} & \textbf{0.0052$\pm$0.0007} & \textbf{34.84$\pm$1.38} & \textbf{0.9596$\pm$0.0081} & \textbf{0.0091$\pm$0.0017} \\
        \bottomrule
    \end{tabular}
    \label{tb3}
\end{table*}

Fig. \ref{fig4} illustrates a qualitative comparison of the reconstructed images and their corresponding error maps using various methods on the fastMRI dataset for 4$\times$ acceleration under 1D equispaced mask. The first row displays the reconstruction resluts from different methods, the second row showcases zoomed-in regions of interest, and the third row demonstrates the error maps. The patterns in the error maps represent reconstruction errors; a less intense color on the map (closer to the cool end of the spectrum) signifies superior reconstruction quality. Clearly, the Zero-filling image exhibits aliasing artifacts and lacks anatomical details. Significantly, compared to other methods, our method reconstructs images with minimal visible artifacts and reconstruction discrepancies compared to the ground truth. Specifically, as depicted in the zoomed-in regions of interest shown in the second row, the detail indicated by the green arrow is only recovered by DUN-SA and MC-CDic. Moreover, at the location indicated by the red arrow, MC-CDic reconstruct a texture that originally does not exist, introducing a mistake. These findings verify that our method not only shows remarkable enhancements in all metrics but also reconstructs more textural details.

\subsection{Results on the IXI dataset}
\begin{figure*}[!ht]
\centerline{\includegraphics[width=\textwidth]{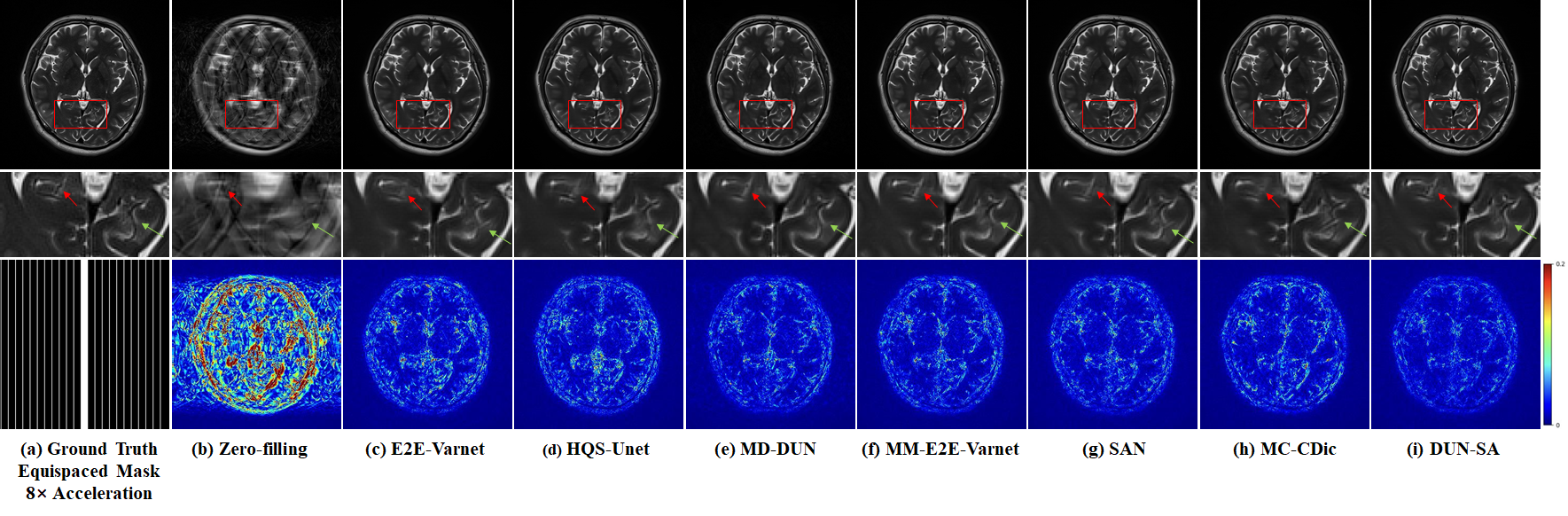}}
\caption{Visual comparison with representative methods for 8$\times$ acceleration under 1D equispaced subsampling mask on the In-house dataset. First row: Reconstructed images by different methods; second row: Zoomed-in region of interest; third row: Equispaced mask of 8$\times$ acceleration and error maps of different methods.}
\label{fig6}
\end{figure*}
\begin{table*}[!ht]
    \centering
    \setlength\tabcolsep{12pt}
    \renewcommand{\arraystretch}{0.85}
    \small
    \caption{Quantitative evaluation of DUN-SA vs. other methods on the BraTs 2018 dataset for 4$\times$ and 8$\times$ acceleration under \textbf{equispaced} and \textbf{random} 1D subsampling masks, where the Reference/Target modalities are T2/FLAIR; T1/FLAIR; T1CE/FLAIR; T1/T1CE; T1CE/T2. Best results are emphasized in \textbf{bold}.}
    \begin{tabular}{@{}l l c c c c@{}}
        \toprule
        \textbf{Ref/Tgt} & \textbf{Methods} & \textbf{Equispaced 4$\times$} & \textbf{Equispaced 8$\times$} & \textbf{Random 4$\times$} & \textbf{Random 8$\times$} \\
        \midrule
        \multirow{8}{*}{\hspace{1.2em}\rotatebox[origin=c]{90}{\textbf{T2/FLAIR}}} 
        & \multirow{2}{*}{MC-CDic} & 44.72 $\pm$ 2.50 & 37.37 $\pm$ 2.19 & 42.09 $\pm$ 2.34 & 37.12 $\pm$ 2.29 \\
        & & 0.9906 $\pm$ 0.0034 & 0.9602 $\pm$ 0.0134 & 0.9821 $\pm$ 0.0059 & 0.9567 $\pm$ 0.0148 \\
        & \multirow{2}{*}{MM-E2E-Varnet} & 45.64 $\pm$ 2.54 & 37.53 $\pm$ 2.25 & 44.14 $\pm$ 2.45 & 37.84 $\pm$ 2.33 \\
        & & 0.9920 $\pm$ 0.0031 & 0.9603 $\pm$ 0.0128 & 0.9886 $\pm$ 0.0042 & 0.9613 $\pm$ 0.0131 \\
        & \multirow{2}{*}{SAN} & 46.03 $\pm$ 2.63 & 38.41 $\pm$ 2.35 & 44.60 $\pm$ 2.48 & 37.88 $\pm$ 2.31 \\
        & & 0.9926 $\pm$ 0.0029 & 0.9645 $\pm$ 0.0125 & 0.9895 $\pm$ 0.0037 & 0.9613 $\pm$ 0.0131 \\
        & \multirow{2}{*}{DUN-SA} & \textbf{48.14 $\pm$ 2.83} & \textbf{39.43 $\pm$ 2.40} & \textbf{46.93 $\pm$ 2.66} & \textbf{39.12 $\pm$ 2.42} \\
        & & \textbf{0.9950 $\pm$ 0.0023} & \textbf{0.9694 $\pm$ 0.0111} & \textbf{0.9932 $\pm$ 0.0027} & \textbf{0.9682 $\pm$ 0.0114} \\
        \midrule
        \multirow{8}{*}{\hspace{1.2em}\rotatebox[origin=c]{90}{\textbf{T1/FLAIR}}} 
        & \multirow{2}{*}{MC-CDic} & 44.83 $\pm$ 2.43 & 37.40 $\pm$ 2.03 & 41.86 $\pm$ 2.14 & 37.10 $\pm$ 2.06 \\
        & & 0.9907 $\pm$ 0.0032 & 0.9598 $\pm$ 0.0112 & 0.9815 $\pm$ 0.0059 & 0.9567 $\pm$ 0.0120 \\
        & \multirow{2}{*}{MM-E2E-Varnet} & 46.03 $\pm$ 2.53 & 38.05 $\pm$ 2.02 & 43.82 $\pm$ 2.19 & 37.66 $\pm$ 2.11 \\
        & & 0.9927 $\pm$ 0.0028 & 0.9627 $\pm$ 0.0126 & 0.9878 $\pm$ 0.0078 & 0.9603 $\pm$ 0.0110 \\
        & \multirow{2}{*}{SAN} & 46.03 $\pm$ 2.63 & 38.09 $\pm$ 2.03 & 44.60 $\pm$ 2.21 & 37.67 $\pm$ 2.10 \\
        & & 0.9926 $\pm$ 0.0029 & 0.9627 $\pm$ 0.0134 & 0.9895 $\pm$ 0.0081 & 0.9603 $\pm$ 0.0109 \\
        & \multirow{2}{*}{DUN-SA} & \textbf{48.34 $\pm$ 2.81} & \textbf{39.03 $\pm$ 2.19} & \textbf{46.51 $\pm$ 2.57} & \textbf{38.81 $\pm$ 2.24} \\
        & & \textbf{0.9952 $\pm$ 0.0021} & \textbf{0.9677 $\pm$ 0.0097} & \textbf{0.9927 $\pm$ 0.0027} & \textbf{0.9666 $\pm$ 0.0102} \\
        \midrule
        \multirow{8}{*}{\hspace{1.2em}\rotatebox[origin=c]{90}{\textbf{T1CE/FLAIR}}} 
        & \multirow{2}{*}{MC-CDic} & 43.17 $\pm$ 2.24 & 37.31 $\pm$ 2.28 & 40.96 $\pm$ 2.26 & 36.94 $\pm$ 2.26 \\
        & & 0.9877 $\pm$ 0.0038 & 0.9582 $\pm$ 0.0129 & 0.9789 $\pm$ 0.0136 & 0.9552 $\pm$ 0.0136 \\
        & \multirow{2}{*}{MM-E2E-Varnet} & 46.22 $\pm$ 2.62 & 37.87 $\pm$ 2.15 & 44.16 $\pm$ 2.44 & 37.53 $\pm$ 2.15 \\
        & & 0.9930 $\pm$ 0.0028 & 0.9611 $\pm$ 0.0149 & 0.9887 $\pm$ 0.0037 & 0.9587 $\pm$ 0.0120 \\
        & \multirow{2}{*}{SAN} & 46.37 $\pm$ 2.63 & 38.02 $\pm$ 2.27 & 44.04 $\pm$ 2.42 & 37.60 $\pm$ 2.25 \\
        & & 0.9932 $\pm$ 0.0028 & 0.9618 $\pm$ 0.0114 & 0.9882 $\pm$ 0.0040 & 0.9593 $\pm$ 0.0119 \\
        & \multirow{2}{*}{DUN-SA} & \textbf{48.19 $\pm$ 2.84} & \textbf{38.98 $\pm$ 2.32} & \textbf{46.50 $\pm$ 2.65} & \textbf{38.70 $\pm$ 2.38} \\
        & & \textbf{0.9951 $\pm$ 0.0022} & \textbf{0.9670 $\pm$ 0.0104} & \textbf{0.9926 $\pm$ 0.0028} & \textbf{0.9655 $\pm$ 0.0110} \\
        \midrule
        \multirow{8}{*}{\hspace{1.2em}\rotatebox[origin=c]{90}{\textbf{T1/T1CE}}} 
        & \multirow{2}{*}{MC-CDic} & 48.60 $\pm$ 2.52 & 41.71 $\pm$ 2.39 & 46.10 $\pm$ 2.40 & 41.51 $\pm$ 2.35 \\
        & & 0.9935 $\pm$ 0.0029 & 0.9764 $\pm$ 0.0109 & 0.9894 $\pm$ 0.0047 & 0.9760 $\pm$ 0.0110 \\
        & \multirow{2}{*}{MM-E2E-Varnet} & 50.18 $\pm$ 1.98 & 41.51 $\pm$ 2.27 & 48.47 $\pm$ 2.32 & 42.05 $\pm$ 2.44 \\
        & & 0.9959 $\pm$ 0.0023 & 0.9769 $\pm$ 0.0104 & 0.9933 $\pm$ 0.0033 & 0.9778 $\pm$ 0.0103 \\
        & \multirow{2}{*}{SAN} & 50.22 $\pm$ 2.66 & 42.24 $\pm$ 2.41 & 48.70 $\pm$ 2.62 & 42.13 $\pm$ 2.44 \\
        & & 0.9960 $\pm$ 0.0022 & 0.9788 $\pm$ 0.0101 & 0.9936 $\pm$ 0.0043 & 0.9782 $\pm$ 0.0102 \\
        & \multirow{2}{*}{DUN-SA} & \textbf{52.25 $\pm$ 3.05} & \textbf{43.47 $\pm$ 2.59} & \textbf{51.35 $\pm$ 2.97} & \textbf{43.22 $\pm$ 2.60} \\
        & & \textbf{0.9971 $\pm$ 0.0018} & \textbf{0.9820 $\pm$ 0.0093} & \textbf{0.9960 $\pm$ 0.0023} & \textbf{0.9809 $\pm$ 0.0095} \\
        \midrule
        \multirow{8}{*}{\hspace{1.2em}\rotatebox[origin=c]{90}{\textbf{T1CE/T2}}} 
        & \multirow{2}{*}{MC-CDic} & 44.19 $\pm$ 1.77 & 37.29 $\pm$ 2.17 & 41.36 $\pm$ 1.69 & 36.17 $\pm$ 1.83 \\
        & & 0.9930 $\pm$ 0.0023 & 0.9711 $\pm$ 0.0116 & 0.9856 $\pm$ 0.0044 & 0.9651 $\pm$ 0.0115 \\
        & \multirow{2}{*}{MM-E2E-Varnet} & 46.61 $\pm$ 1.99 & 37.43 $\pm$ 1.84 & 44.54 $\pm$ 1.84 & 37.32 $\pm$ 1.92 \\
        & & 0.9958 $\pm$ 0.0016 & 0.9713 $\pm$ 0.0130 & 0.9921 $\pm$ 0.0026 & 0.9709 $\pm$ 0.0104 \\
        & \multirow{2}{*}{SAN} & 46.73 $\pm$ 2.00 & 37.73 $\pm$ 1.86 & 44.78 $\pm$ 1.88 & 37.38 $\pm$ 1.86 \\
        & & 0.9959 $\pm$ 0.0015 & 0.9726 $\pm$ 0.0101 & 0.9924 $\pm$ 0.0026 & 0.9711 $\pm$ 0.0104 \\
        & \multirow{2}{*}{DUN-SA} & \textbf{49.22 $\pm$ 2.22} & \textbf{39.29 $\pm$ 2.40} & \textbf{48.07 $\pm$ 2.37} & \textbf{39.03 $\pm$ 1.98} \\
        & & \textbf{0.9974 $\pm$ 0.0012} & \textbf{0.9782 $\pm$ 0.0073} & \textbf{0.9958 $\pm$ 0.0017} & \textbf{0.9774 $\pm$ 0.0092} \\
        \bottomrule
    \end{tabular}
    \label{tb4}
\end{table*}

\begin{figure*}[!ht]
\centerline{\includegraphics[width=\textwidth]{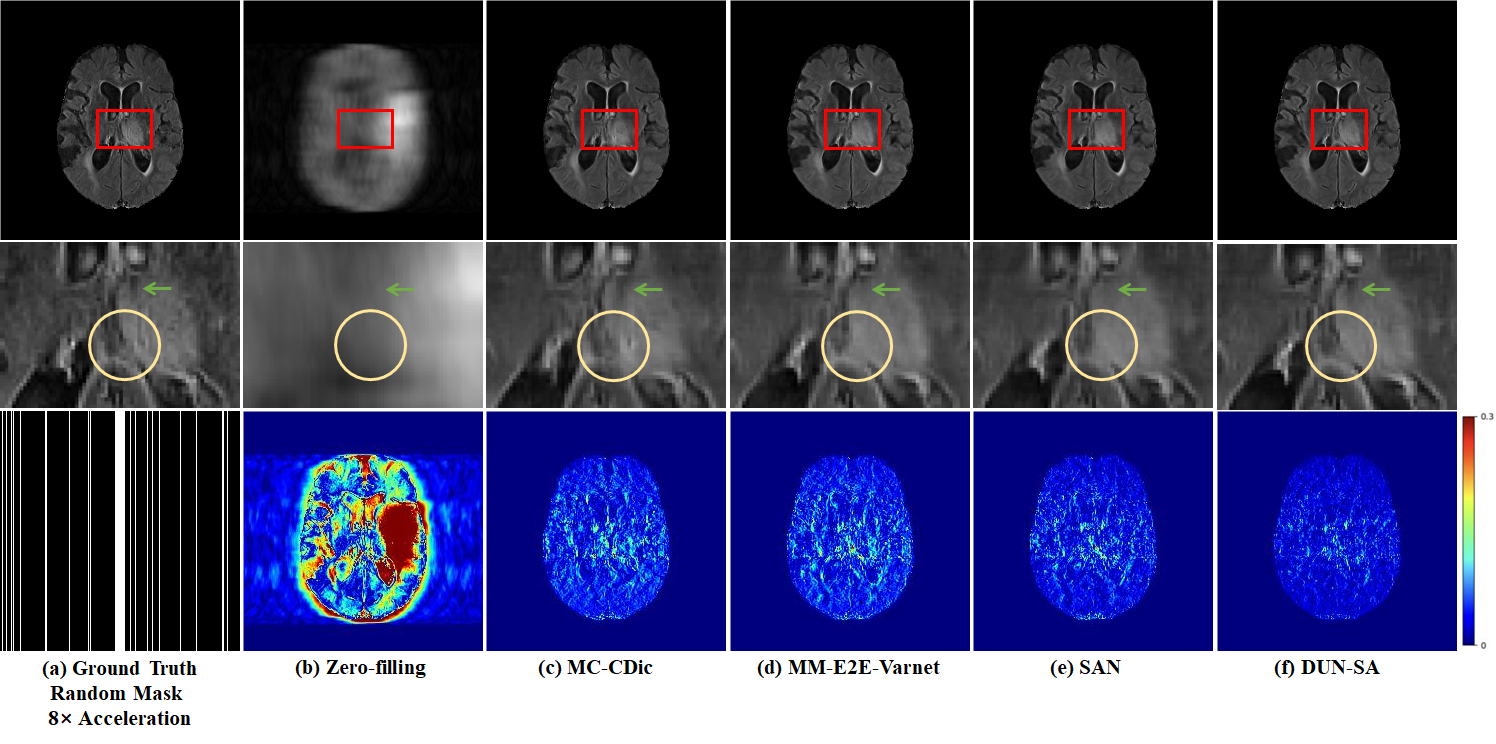}}
\caption{Visual comparison with representative methods for 8$\times$ acceleration under 1D random subsampling mask on the BraTs 2018 dataset. First row: Reconstructed images by different methods; second row: Zoomed-in region of interest; third row: Random mask of 8$\times$ acceleration and error maps of different methods.}
\label{fig7}
\end{figure*}
We validate the effectiveness of our method on a larger dataset, IXI. In this part of the experiments, the fully sampled PD-weighted image is used as reference modality image to assist the reconstruction of the paired under-sampled T2-weighted image.

Table \ref{tb2} presents the quantitative evaluations on the IXI dataset for 1D equispaced and random masks under 4× and 8× acceleration. From the metrics, it can be observed that the reference modality can significantly enhance the reconstruction performance, and the spatial alignment further improves all metrics, which is consistent with the experimental results on the fastMRI dataset. Our method once again brings noticeable improvements for every metric across all settings compared to other methods. For instance, compared to the second-best values, under equispaced 1D subsampling masks of 4$\times$ acceleration, DUN-SA improves the PSNR value from SAN's 46.17db to 47.25db. Meanwhile, under random 1D subsampling masks of 8$\times$ acceleration, DUN-SA enhances SAN's SSIM value from 0.9829 to 0.9846 and reduces the MAE value from 0.0050 to 0.0047.

Fig. \ref{fig5} illustrates a qualitative comparison of the reconstructed images of different methods and their corresponding error maps on the IXI dataset for 8$\times$ acceleration under equispaced 1D subsampling mask. The content displayed is consistent with Fig. \ref{fig4}. Specifically, from the error maps, it's evident that our method yields the minimal error. Observing the zoomed-in region of interest, in the area indicated by the green arrow, only DUN-SA successfully reconstructs the relevant texture details. In contrast, in the region pointed out by the red arrow, other methods either introduce excessive noise or fail to reconstruct the corresponding texture, but the proposed DUN-SA can reconstruct relatively clear details without being significantly affected by noise. These results suggest that our approach is capable of producing high-quality reconstruction results that present tissues with fewer artifacts and noise compared to other methods.

\subsection{Results on the In-house dataset}

In this section, we conduct experiments on the In-house dataset, where the T1-weighted image is used as the reference modality to assist in the reconstruction of the T2-weighted image.

Quantitative results on the In-house dataset for 1D equispaced and random masks under 4× and 8× acceleration are presented in Table \ref{tb3}, showing that the proposed DUN-SA outperforms other methods in achieving the best results for every metric across all settings.

Qualitative results on the In-house dataset for 1D equispaced mask under 8× acceleration are depicted in Fig. \ref{fig6}. Similarly, by comparing the reconstructed images of different methods and their corresponding error maps, it can be observed that the proposed DUN-SA obtains the best results, reconstructing more details. Specifically, in zoomed-in regions of interest shown in the second row, the details indicated by the red arrow and green row are only reconstructed by the proposed DUN-SA. This in line with the performance improvement observed in the aforementioned two datasets and indicates the promising generalizability of DUN-SA.

\subsection{Results on the BraTS 2018 dataset}
\begin{figure*}[!ht]
\centerline{\includegraphics[width=\textwidth]{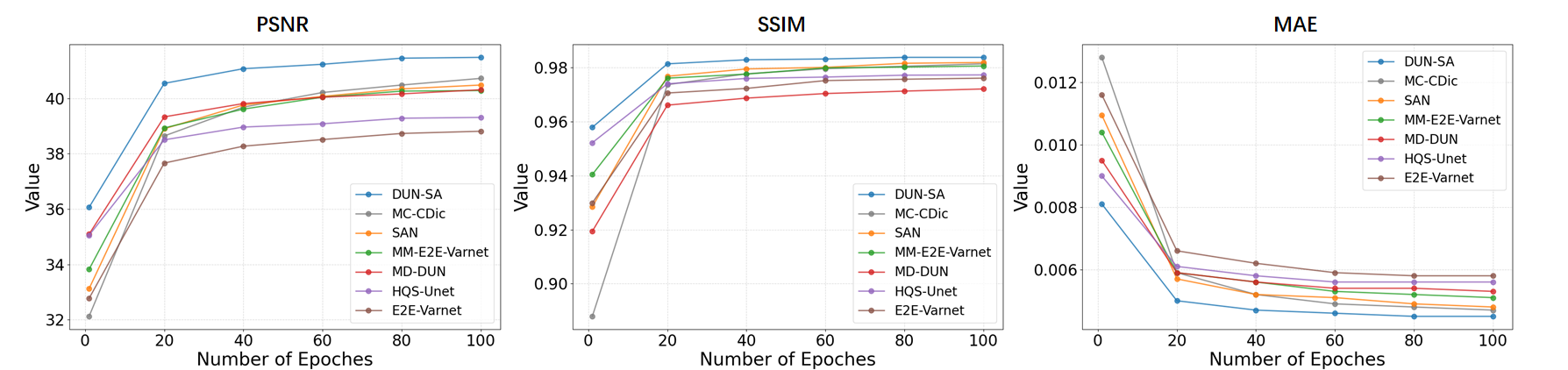}}
\caption{Comparison on the learning trajectories of different models on the fastMRI dataset for 4$\times$ acceleration under equispaced mask.}
\label{fig8}
\end{figure*}
\begin{figure*}[!ht]
\centerline{\includegraphics[width=\textwidth]{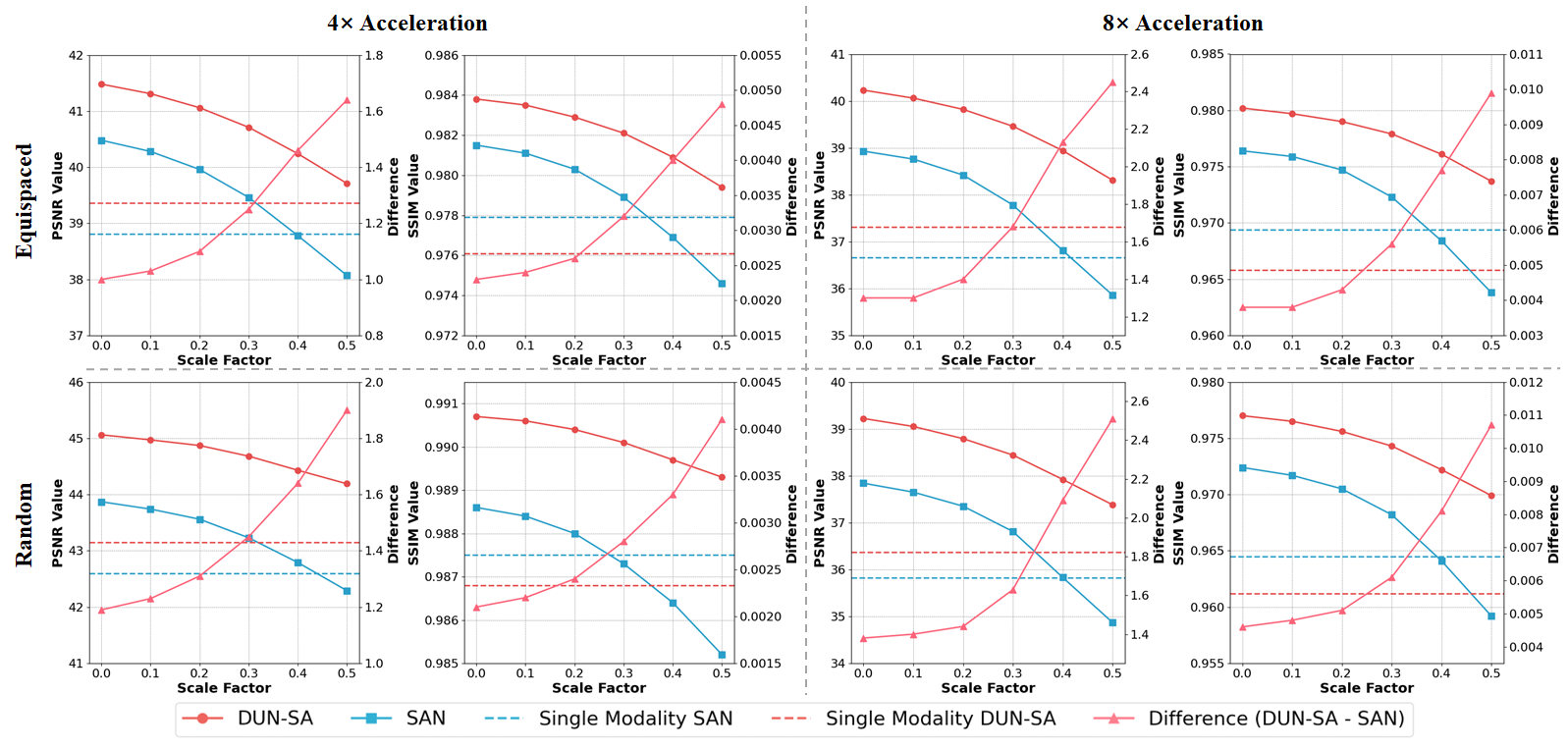}}
\caption{Quantitative comparison of multi-modal MRI reconstruction on the fastMRI dataset with different scales of simulated spatial misalignment. Left y-axes are for reconstruction performances (“DUN-SA”, “SAN”, “Single Modality SAN” and “Single Modality DUN-SA”) while the right y-axes are for the “Difference” between “DUN-SA” and “SAN”.}
\label{fig9}
\end{figure*}
In this section, we conduct experiments on the BraTS 2018 dataset to assess model performance with greater variability in protocols and pathologies. Specifically, we test the cases where the Reference/Target modalities are T2/FLAIR; T1/FLAIR; T1CE/FLAIR; T1/T1CE; T1CE/T2. Moreover, we compare the selected best-performing four multi-modal methods across the three mentioned datasets.

Quantitative results on the BraTS 2018 dataset for 1D equispaced and random masks under 4$\times$ and 8$\times$ acceleration are presented in Table \ref{tb4}. The results indicate that the proposed DUN-SA maintains its high performance when facing variability in protocols and pathologies, further illustrating the generalizability of DUN-SA.

Qualitative results on the BraTS 2018 dataset for 1D random mask under 8× acceleration are depicted in Fig. \ref{fig7}. It can be observed from the Fig. 7 that, even when reconstructing MRI images with tumor, DUN-SA still recovers the most details, maintaining the best performance. Especially, in the zoomed-in regions of interest in the second row, we demonstrate the reconstruction performance around the tumor. In the areas indicated by green arrows, only DUN-SA and MC-CDic are able to reconstruct such signals; however, the signals reconstructed by MC-CDic are comparatively vaguer. Furthermore, within the yellow circles, DUN-SA recovered more precise and clearer edge details.

\section{Discussion}
In this section, we first compare the convergence performance of our proposed DUN-SA method with other methods during the training process. Subsequently, we carry out experiments to assess its performance under varying degrees of misalignment. Furthermore, we compare its spatial alignment capabilities with those of alternative methods. We evaluate the adaptability of the reconstruction performance in scenarios where the reference data is imperfect. Ablation studies are conducted to determine the optimal number of stages and to validate the advantages of the key components. We employ DUN-SA to execute a model verification experiment aimed at elucidating the operational mechanism behind the network modules. Finally, we analyze the model complexity.

\subsection{Convergence of DUN-SA vs. other methods}
In this section, we compare the convergence performance of the proposed DUN-SA with other methods by recording and comparing the reconstruction performance of each method throughout the training process. Specifically, we record the PSNR, SSIM, and MAE metrics during the training process on the fastMRI dataset for 4$\times$ acceleration under equispaced mask. The curves in Fig. \ref{fig8} illustrate that the convergence performance of DUN-SA is superior compared to the other methods because it consistently maintains higher PSNR and SSIM values and a lower MAE throughout the entire training process.
\subsection{Performance evaluation under different scales of misalignment between modalities}
\begin{figure*}[!ht]
\centerline{\includegraphics[width=\textwidth]{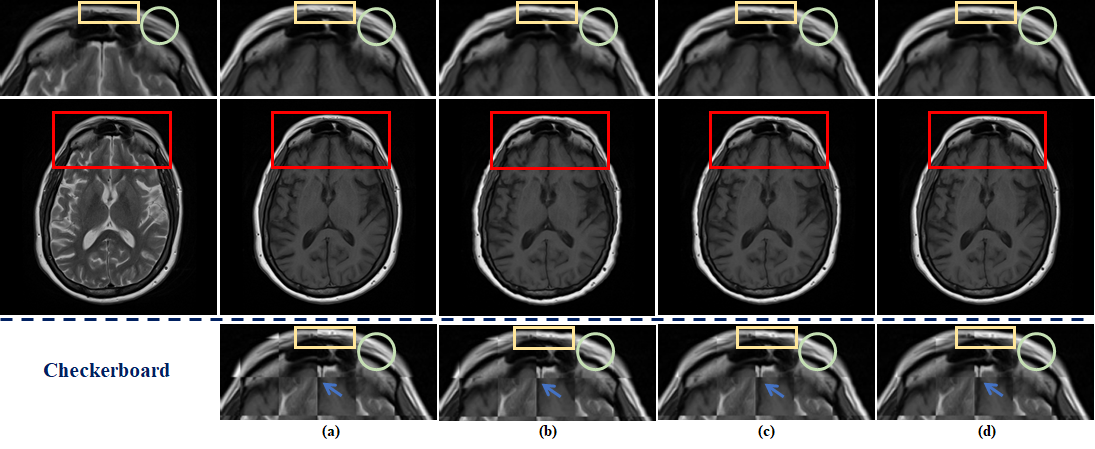}}
\caption{The visualization of the effects of spatial alignment on the fastMRI dataset. (a) shows the original fully-sampled T1 image. (b) represents the results of aligning an under-sampled T2 image with a fully-sampled T1 image by traditional method. (c) depicts the results of integrating traditional spatial alignment with reconstruction for joint optimization. (d) displays the result of DUN-SA. Details are shown in the first row:  zoomed-in view of aligned T1 images and third row: zoomed-in views of checkerboard visualizations.}
\label{fig10}
\end{figure*}

\begin{figure*}[!ht]
\centerline{\includegraphics[width=\textwidth]{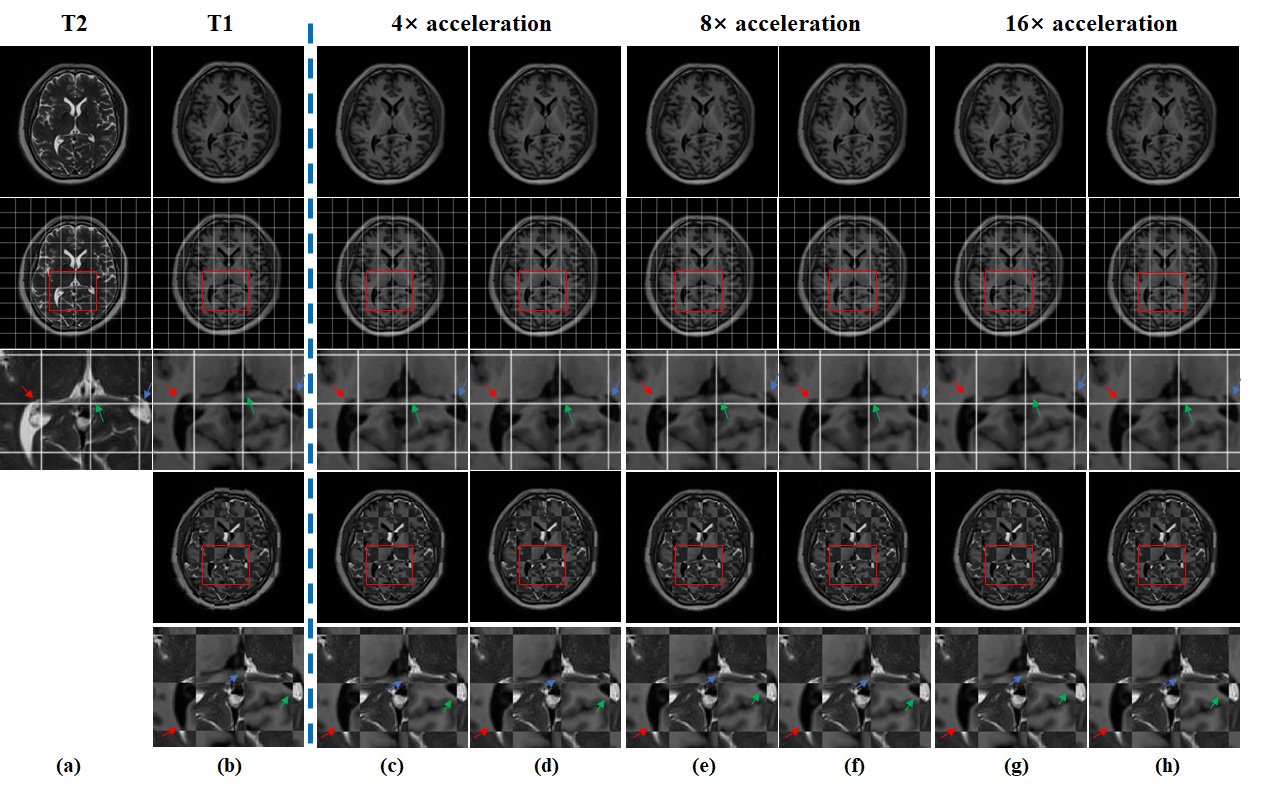}}
\caption{The visualization of the effects of spatial alignment on the In-house dataset. (a), (b) represent fully-sampled T2-weighted image and fully-sampled T1-weighted image, respectively. (c), (e), and (g) depict T1-weighted images aligned using SAN under different acceleration factors, while (d), (f), and (h) display T1-weighted images aligned using DUN-SA under different acceleration factors. In the second row, a grid is used to facilitate observation of the spatial position of each aforementioned image, with zoomed-in views presented in the third row. In the fourth row, checkerboard visualizations are employed to demonstrate the misalignment between T2-weighted image and T1-weighted image/aligned T1-weighted image, and the last row magnifies the corresponding areas to display the details more clearly.}
\label{fig11}
\end{figure*}
To demonstrate the superior robustness of the proposed DUN-SA to misalignment between modalities, we compare it with SAN, another reconstruction method that considers misalignment. We perform different scales of spatial misalignment on the reference modality and then test using the parameters in Section \ref{efd}. This experiment is conducted on the fastMRI dataset for 4$\times$ and 8$\times$ acceleration under both equispaced and random 1D subsampling masks. The method for simulating spatial misalignment is consistent with the approach in SAN; specifically, we employ random rotations within the range of \( [-0.01 \pi \sigma, 0.01 \pi \sigma] \), translations between \( [-0.05 N \sigma, 0.05 N \sigma] \), and displacement field bicubically interpolated from \( 9 \times 9 \) control-points (with displacements uniformly sampled within \( [-0.02 N \sigma, 0.02 N \sigma] \) in both directions. Here, \( N \) represents the size of the MR images and \( \sigma \) indicates the factor controlling the degree of spatial misalignment). Fig. \ref{fig9} compares the reconstruction performance of the two methods under different degrees of spatial misalignment. It's noteworthy that as the scale of spatial misalignment increases, the performance of both methods declines. However, DUN-SA degrades more slowly, whereas the performance of SAN decreases relatively faster, leading to a gradually increasing difference between the two methods. This highlights the greater robustness of the proposed DUN-SA to spatial misalignment.

\subsection{Comparison of spatial alignment performance}
In this section, we compare our spatial alignment result with an effective and widely-used traditional spatial alignment method \citep{sandkuhler2018airlab}. In Fig. \ref{fig10}, (a) shows the original fully-sampled T1 image, demonstrating the misalignment between modalities; (b) represents the results of aligning an under-sampled T2 image with a fully-sampled T1 image by traditional method, illustrating the limitations of traditional method in this scenario; utilizing traditional methods to address the spatial alignment task and integrating it with the reconstruction task for joint optimization can further improve the spitial alignment performance as depicted in (c). However, the outcome is still inferior to that achieved using DUN-SA, as displayed in (d). Specifically, although the traditional method can achieve general alignment, as indicated by the blue arrow and yellow rectangle in the checkerboard patterns, it results in some local deformations due to the artifacts caused by undersampling of the target modality image and the inherent difficulties of the cross-modal spatial alignment task. These deformations are clearly displayed in the first row, highlighted by the yellow rectangle and green circle. In contrast, our method not only achieves general alignment but also preserves the local structures.

Additionally, we evaluate the spatial alignment performance of the proposed DUN-SA under different acceleration factors and compare it with SAN. In Fig. \ref{fig11}, (a) and (b) represent fully-sampled T2-weighted image and fully-sampled T1-weighted image, respectively. (c), (e), and (g) depict T1-weighted images aligned using SAN under different acceleration factors, while (d), (f), and (h) display T1-weighted images aligned using DUN-SA under different acceleration factors. We use a grid to facilitate the observation of spatial positions and employ a checkerboard to visualize misalignment. We find that both DUN-SA and SAN exhibit good spatial alignment performance at 4$\times$ acceleration. However, as the acceleration factor increases, SAN gradually fails to align T1 and T2 well, whereas DUN-SA is minimally affected. For instance, we can observe in the zoomed-in views of the third and fifth rows, at the locations indicated by the red, blue, and green arrows, misalignment still exists in (e) and (g), but this misalignment has been alleviated in (f) and (h). This demonstrates that DUN-SA, by iteratively solving the proposed model, appropriately integrates the spatial alignment task into the reconstruction process and achieves more precise spatial alignment results under high acceleration factors.

\subsection{Adaption to scenarios with imperfect reference data}
\begin{table}[!ht]
    \centering
    \small
    \renewcommand{\arraystretch}{0.89}
    \setlength\tabcolsep{8pt}
    \caption{Evaluation of adaptation to scenarios with imperfect reference data. US-Half/Rec-Half (US-All/Rec-all) indicates that half (all) reference modality data to be under-sampled/reconstructed.}
    \label{tb5}
    \begin{tabular}{cccc}
        \toprule
        & \textbf{Methods} & \textbf{Equispaced 4$\times$} & \textbf{Equispaced 8$\times$} \\
        \midrule
        \multirow{8}{*}{\rotatebox[origin=c]{90}{\textbf{US-Half}}} 
        & \multirow{2}{*}{MC-CDic} & 37.28 $\pm$ 2.34 & 34.06 $\pm$ 3.09 \\
        & & 0.9644 $\pm$ 0.0128 & 0.9398 $\pm$ 0.0293 \\
        & \multirow{2}{*}{MM-E2E-Varnet} & 38.73 $\pm$ 1.93 & 36.95 $\pm$ 2.07 \\
        & & 0.9758 $\pm$ 0.0077 & 0.9674 $\pm$ 0.0106 \\
        & \multirow{2}{*}{SAN} & 38.64 $\pm$ 2.02 & 36.77 $\pm$ 2.22 \\
        & & 0.9755 $\pm$ 0.0079 & 0.9663 $\pm$ 0.0115 \\
        & \multirow{2}{*}{DUN-SA} & \textbf{38.78 $\pm$ 2.03} & \textbf{37.02 $\pm$ 2.19} \\
        & & \textbf{0.9760 $\pm$ 0.0080} & \textbf{0.9675 $\pm$ 0.0112} \\
        \midrule
        \multirow{8}{*}{\rotatebox[origin=c]{90}{\textbf{US-All}}}
        & \multirow{2}{*}{MC-CDic} & 38.23 $\pm$ 1.66 & 35.66 $\pm$ 1.70 \\
        & & 0.9719 $\pm$ 0.0077 & 0.9573 $\pm$ 0.0118 \\
        & \multirow{2}{*}{MM-E2E-Varnet} & 38.99 $\pm$ 1.62 & 36.79 $\pm$ 1.68 \\
        & & 0.9767 $\pm$ 0.0066 & 0.9667 $\pm$ 0.0087 \\
        & \multirow{2}{*}{SAN} & 39.00 $\pm$ 1.62 & 37.04 $\pm$ 1.67 \\
        & & 0.9762 $\pm$ 0.0065 & 0.9679 $\pm$ 0.0085 \\
        & \multirow{2}{*}{DUN-SA} & \textbf{39.87 $\pm$ 1.60} & \textbf{38.23 $\pm$ 1.66} \\
        & & \textbf{0.9793 $\pm$ 0.0063} & \textbf{0.9724 $\pm$ 0.0082} \\
        \midrule
        \multirow{8}{*}{\rotatebox[origin=c]{90}{\textbf{Rec-Half}}} 
        & \multirow{2}{*}{MC-CDic} & 39.40 $\pm$ 1.79 & 36.88 $\pm$ 1.94 \\
        & & 0.9772 $\pm$ 0.0073 & 0.9660 $\pm$ 0.0112 \\
        & \multirow{2}{*}{MM-E2E-Varnet} & 39.71 $\pm$ 1.80 & 37.79 $\pm$ 1.90 \\
        & & 0.9791 $\pm$ 0.0068 & 0.9714 $\pm$ 0.0090 \\
        & \multirow{2}{*}{SAN} & 39.88 $\pm$ 1.79 & 38.21 $\pm$ 1.92 \\
        & & 0.9796 $\pm$ 0.0066 & 0.9729 $\pm$ 0.0086 \\
        & \multirow{2}{*}{DUN-SA} & \textbf{40.74 $\pm$ 1.85} & \textbf{39.39 $\pm$ 2.02} \\
        & & \textbf{0.9818 $\pm$ 0.0062} & \textbf{0.9771 $\pm$ 0.0079} \\
        \midrule
        \multirow{8}{*}{\rotatebox[origin=c]{90}{\textbf{Rec-All}}}
        & \multirow{2}{*}{MC-CDic} & 38.88 $\pm$ 1.70 & 36.39 $\pm$ 1.80 \\
        & & 0.9740 $\pm$ 0.0071 & 0.9630 $\pm$ 0.0104 \\
        & \multirow{2}{*}{MM-E2E-Varnet} & 39.22 $\pm$ 1.70 & 37.36 $\pm$ 1.77 \\
        & & 0.9774 $\pm$ 0.0067 & 0.9693 $\pm$ 0.0087 \\
        & \multirow{2}{*}{SAN} & 39.32 $\pm$ 1.70 & 37.51 $\pm$ 1.81 \\
        & & 0.9777 $\pm$ 0.0065 & 0.9699 $\pm$ 0.0085 \\
        & \multirow{2}{*}{DUN-SA} & \textbf{40.19 $\pm$ 1.73} & \textbf{38.64 $\pm$ 1.81} \\
        & & \textbf{0.9803 $\pm$ 0.0061} & \textbf{0.9744 $\pm$ 0.0077} \\
        \bottomrule
    \end{tabular}
\end{table}

\begin{figure*}[!ht]
\centerline{\includegraphics[width=\textwidth]{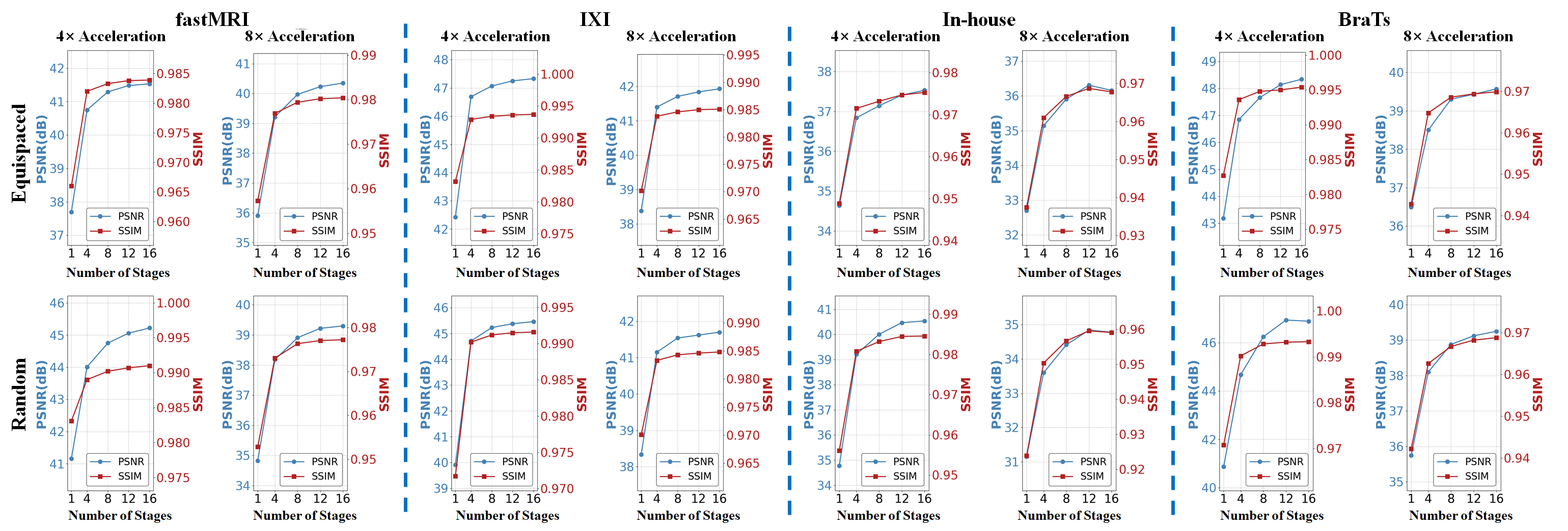}}
\caption{The PSNR and SSIM curves on the fastMRI, IXI, In-house and BraTs datasets with different numbers of stages k.}
\label{fig12}
\end{figure*}

\begin{figure*}[!ht]
\centerline{\includegraphics[width=\textwidth]{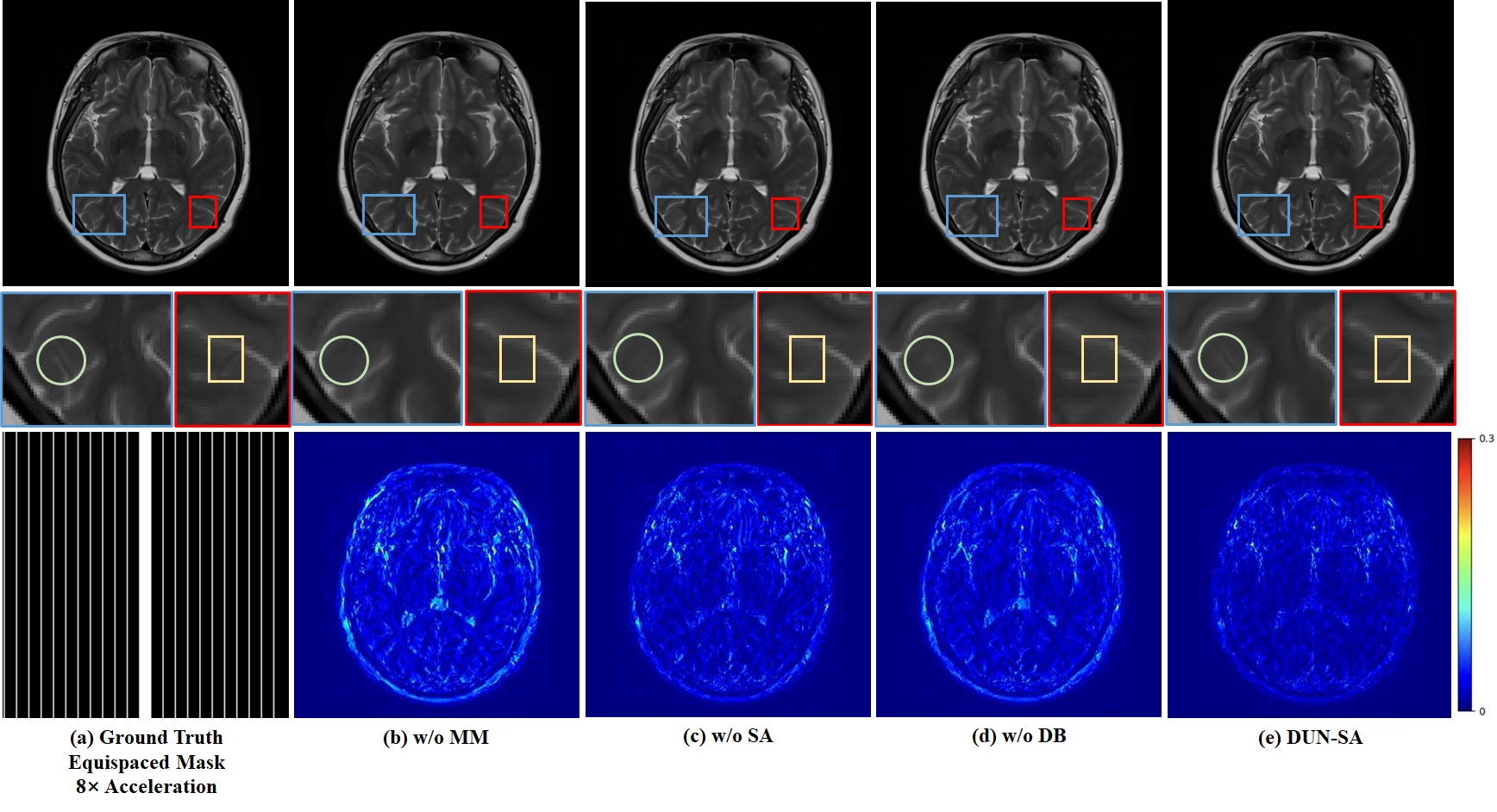}}
\caption{Visual comparison with effect of each component for 8$\times$ acceleration under 1D equispaced subsampling mask on the fastMRI dataset. First row: Reconstructed images by different methods; second row: Zoomed-in region of interest; third row: Equispaced mask of 8$\times$ acceleration and error maps of different methods.}
\label{fig13}
\end{figure*}
In this section, we assess the adaptability of the reconstruction performance to scenarios with imperfect reference data. Specifically, during training, we set the reference modality (either all or half) to be under-sampled for both training and testing. From Table \ref{tb5}, we find that within the paired data, if the reference modality images are under-sampled, the reconstruction quality of the target modality images drops significantly. To mitigate this issue, we explore a method to enhance the flexibility and utility of our model. We train a single-modality reconstruction network (refer to "w/o MM" in Section \ref{ekc}) for the reference modality, first reconstructing the under-sampled reference modality images and then using the reconstructed reference modality to assist in the reconstruction of the target modality. From the experimental results, all multi-modal methods have shown improvement. It is noteworthy that DUN-SA outperforms other multi-modal reconstruction methods across a range of reference data quality, from low to high. This demonstrates that DUN-SA also exhibits superior performance in scenarios with imperfect reference data.

\subsection{Ablation Study}
\begin{table*}[!ht]
\centering
\renewcommand{\arraystretch}{0.89}
\caption{Effect of each component on the performance of DUN-SA on the fastMRI, IXI, In-house and BraTs datasets for 4$\times$ and 8$\times$ acceleration under equispaced and random subsampling masks, measured in PSNR and SSIM.}
\label{tb6}
\setlength\tabcolsep{12pt}
\small
\begin{tabular}{l *{5}{c}}
\toprule
&\textbf{Methods} & \textbf{Equispaced 4$\times$} & \textbf{Equispaced 8$\times$} & \textbf{Random 4$\times$} & \textbf{Random 8$\times$} \\
\midrule
\multirow{8}{*}{\rotatebox[origin=c]{90}{\textbf{fastMRI}}} 
&\multirow{2}{*}{w/o MM} & 39.36 $\pm$ 1.67 & 37.32 $\pm$ 1.76 & 43.35 $\pm$ 1.97 & 36.37 $\pm$ 1.77 \\
&  & 0.9779 $\pm$ 0.0064 & 0.9694 $\pm$ 0.0082 & 0.9875 $\pm$ 0.0039 & 0.9645 $\pm$ 0.0096 \\
&\multirow{2}{*}{w/o SA}  & 41.07 $\pm$ 1.80 & 39.75 $\pm$ 1.95 & 44.68 $\pm$ 2.01 & 38.65 $\pm$ 1.98 \\
&  & 0.9829 $\pm$ 0.0063 & 0.9787 $\pm$ 0.0078 & 0.9899 $\pm$ 0.0039 & 0.9755 $\pm$ 0.0090 \\
&\multirow{2}{*}{w/o DB}  & 41.16 $\pm$ 1.86 & 39.86 $\pm$ 1.88 & 44.70 $\pm$ 1.95 & 38.75 $\pm$ 2.02 \\
&  & 0.9831 $\pm$ 0.0062 & 0.9791 $\pm$ 0.0075 & 0.9900 $\pm$ 0.0041 & 0.9758 $\pm$ 0.0092 \\
&\multirow{2}{*}{DUN-SA}  & \textbf{41.48 $\pm$ 1.84} & \textbf{40.23 $\pm$ 1.88} & \textbf{45.06 $\pm$ 1.85} & \textbf{39.23 $\pm$ 1.84} \\
&  & \textbf{0.9838 $\pm$ 0.0060} & \textbf{0.9802 $\pm$ 0.0072} & \textbf{0.9907 $\pm$ 0.0029} & \textbf{0.9770 $\pm$ 0.0075} \\
\midrule
\multirow{8}{*}{\rotatebox[origin=c]{90}{\textbf{IXI}}}
&\multirow{2}{*}{w/o MM}  & 42.54 $\pm$ 2.32 & 35.23 $\pm$ 2.09 & 40.36 $\pm$ 2.17 & 34.35 $\pm$ 2.05 \\
&  & 0.9871 $\pm$ 0.0039 & 0.9571 $\pm$ 0.0099 & 0.9821 $\pm$ 0.0052 & 0.9501 $\pm$ 0.0107 \\
&\multirow{2}{*}{w/o SA}  & 47.12 $\pm$ 2.54 & 41.72 $\pm$ 2.33 & 45.27 $\pm$ 2.62 & 41.47 $\pm$ 2.32 \\
&  & 0.9935 $\pm$ 0.0026 & 0.9848 $\pm$ 0.0053 & 0.9914 $\pm$ 0.0033 & 0.9843 $\pm$ 0.0055 \\
&\multirow{2}{*}{w/o DB}  & 46.94 $\pm$ 2.51 & 41.57 $\pm$ 2.31 & 44.98 $\pm$ 2.59 & 41.33 $\pm$ 2.30 \\
&  & 0.9933 $\pm$ 0.0026 & 0.9846 $\pm$ 0.0053 & 0.9912 $\pm$ 0.0034 & 0.9839 $\pm$ 0.0056 \\
&\multirow{2}{*}{DUN-SA}  & \textbf{47.25 $\pm$ 2.53} & \textbf{41.84 $\pm$ 2.33} & \textbf{45.38 $\pm$ 2.51} & \textbf{41.62 $\pm$ 2.29} \\
&  & \textbf{0.9936 $\pm$ 0.0026} & \textbf{0.9850 $\pm$ 0.0053} & \textbf{0.9915 $\pm$ 0.0032} & \textbf{0.9846 $\pm$ 0.0054} \\
\midrule
\multirow{8}{*}{\rotatebox[origin=c]{90}{\textbf{In-house}}} 
&\multirow{2}{*}{w/o MM}  & 35.74 $\pm$ 0.76 & 34.02 $\pm$ 0.77 & 37.78 $\pm$ 0.79 & 32.62 $\pm$ 0.74 \\
&  & 0.9658 $\pm$ 0.0029 & 0.9531 $\pm$ 0.0039 & 0.9753 $\pm$ 0.0017 & 0.9458 $\pm$ 0.0042 \\
&\multirow{2}{*}{w/o SA}  & 37.06 $\pm$ 1.22 & 35.84 $\pm$ 1.31 & 40.02 $\pm$ 1.11 & 34.40 $\pm$ 1.34 \\
&  & 0.9727 $\pm$ 0.0052 & 0.9663 $\pm$ 0.0069 & 0.9834 $\pm$ 0.0025 & 0.9563 $\pm$ 0.0082 \\
&\multirow{2}{*}{w/o DB}  & 37.20 $\pm$ 1.17 & 35.93 $\pm$ 1.37 & 39.94 $\pm$ 1.09 & 34.53 $\pm$ 1.16 \\
&  & 0.9739 $\pm$ 0.0044 & 0.9671 $\pm$ 0.0067 & 0.9832 $\pm$ 0.0024 & 0.9584 $\pm$ 0.0061 \\
&\multirow{2}{*}{DUN-SA}  & \textbf{37.41 $\pm$ 1.08} & \textbf{36.31 $\pm$ 1.41} & \textbf{40.47 $\pm$ 1.09} & \textbf{34.84 $\pm$ 1.38} \\
&  & \textbf{0.9747 $\pm$ 0.0039} & \textbf{0.9687 $\pm$ 0.0070} & \textbf{0.9845 $\pm$ 0.0024} & \textbf{0.9596 $\pm$ 0.0081} \\
\midrule
\multirow{8}{*}{\rotatebox[origin=c]{90}{\textbf{BraTs 2018}}} 
&\multirow{2}{*}{w/o MM}  & 47.19 $\pm$ 2.77 & 37.55 $\pm$ 2.23 & 45.32 $\pm$ 2.38 & 37.16 $\pm$ 2.26 \\
&  & 0.9934 $\pm$ 0.0026 & 0.9605 $\pm$ 0.0112 & 0.9921 $\pm$ 0.0030 & 0.9569 $\pm$ 0.0146 \\
&\multirow{2}{*}{w/o SA}  & 47.92 $\pm$ 2.83 & 39.16 $\pm$ 2.51 & 46.55 $\pm$ 2.51 & 38.91 $\pm$ 2.42 \\
&  & 0.9947 $\pm$ 0.0024 & 0.9682 $\pm$ 0.0114 & 0.9930 $\pm$ 0.0028 & 0.9667 $\pm$ 0.0118 \\
&\multirow{2}{*}{w/o DB}  & 47.80 $\pm$ 2.78 & 38.98 $\pm$ 2.35 & 46.53 $\pm$ 2.53 & 38.74 $\pm$ 2.36 \\
&  & 0.9945 $\pm$ 0.0022 & 0.9676 $\pm$ 0.0115 & 0.9929 $\pm$ 0.0028 & 0.9655 $\pm$ 0.0110 \\
&\multirow{2}{*}{DUN-SA}  & \textbf{48.14 $\pm$ 2.83} & \textbf{39.43 $\pm$ 2.40} & \textbf{46.93 $\pm$ 2.66} & \textbf{39.12 $\pm$ 2.42} \\
&  & \textbf{0.9950 $\pm$ 0.0023} & \textbf{0.9694 $\pm$ 0.0111} & \textbf{0.9932 $\pm$ 0.0027} & \textbf{0.9682 $\pm$ 0.0114} \\
\bottomrule
\end{tabular}
\end{table*}
To determine the optimal network architecture, we conduct two ablation studies. The first study focus on the number of stages, and the second emphasize each component within the network. The reference and target modalities are configured as follows: T1/T2 for fastMRI dataset; PD/T2 for IXI dataset; T1/T2 for In-house dataset; and T2/FLAIR for BraTs 2018 dataset. 

\subsubsection{Effect of number of stages}
To demonstrate how the number of stages \(k\) affects the reconstruction performance, we carry out a quantitative comparison of DUN-SA under different numbers of stages on four datasets mentioned for $4\times$ and $8\times$ acceleration under equispaced and random subsampling masks. Fig. \ref{fig12} shows the mean SSIM and PSNR values for stages ranging from 1 to 16. We observe that from the first iteration to the twelfth iteration, the two metrics show a obviously progressive improvement. From the thirteenth iteration, the improvements in the two metrics became subtle or began to show a downward trend. Taking both performance and model complexity into consideration, we chose the model with \(k=12\).

\subsubsection{Effect of key components}
\begin{figure*}[!ht]
\centerline{\includegraphics[width=\textwidth]{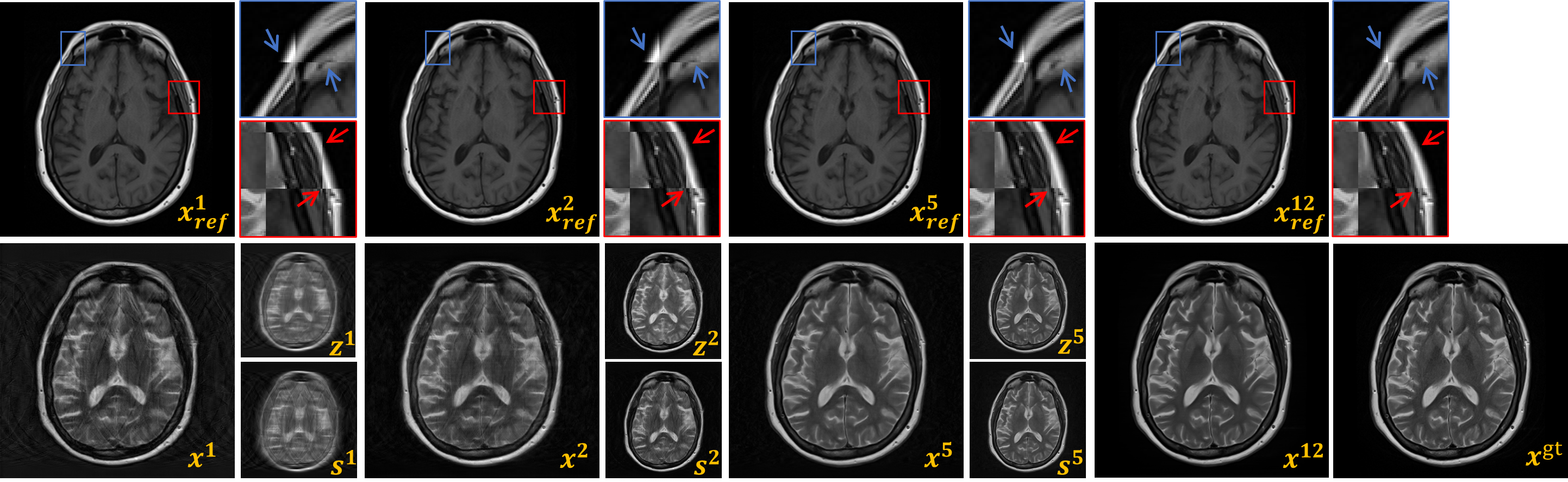}}
\caption{Visualization of Intermediate Results at Stage t: ground truth $x^{gt}$, reconstructed image $x^t$, warped reference image $\mathcal{T}(x_{\text{ref}}, \phi^{t})$ denoted as $x_{ref}^t$, inter-modality prior $z^t$, and intra-modality prior $s^t$.}
\label{fig14}
\end{figure*}
Based on the optimal number of iterations of 12 for the first ablation experiment, we conduct the second ablation experiment on our proposed DUN-SA on fastMRI, IXI, In-house, BraTs 2018 datasets as follows:

\textbf{Aligned Inter-modality Prior Learning Block (AIPLB):} The ablation study for AIPLB means removing the AIPLB module from the network. It's worth noting that, when the AIPLB module is removed, the assistance from the reference modality will also be omitted, and the spatial alignment operator will be simultaneously disregarded. The model degrades into a single-modal MRI reconstruction model. This configuration is termed the proposed DUN-SA without multi-modality (w/o MM).

\textbf{Spatial Alignment Module (SAM):} The ablation study for SAM involves removing the SAM module from the network. In this scenario, the spatial alignment operator is ignored within the network. As a result, the reference modality images are directly concated with intermediate reconstructed target modality images and feed in $\text{ProxNet}_Z$ to learn the inter-modality prior. This configuration is termed the proposed DUN-SA without spatial alignment (w/o SA).

\textbf{Denoising Block (DB):} The ablation study for DB refers to the removal of the DB module from the network, which implies the elimination of the $\text{ProxNet}_S$'s function and the removal of intra-modality prior. This configuration is termed the proposed DUN-SA without denoising block (w/o DB).

Table \ref{tb6} and Fig. \ref{fig13} illustrate quantitative and qualitative comparisons of the reconstructed images from the four aforementioned methods in ablation study. From the results, we can find that (1) w/o MM shows the worst reconstruction results in terms of PSNR and SSIM. Additionally, from the error map, it is evident that this configuration results in the most severe loss of structural details and the most significant discrepancies compared to the ground-truth image. This clearly validates that the reference modality provides extra inter-modality prior to assist in the reconstruction of more structural details as presented in Section \ref{of}. (2) From the results, we can observe that w/o SA performs inferiorly compared to DUN-SA, as indicated by the brighter error map and lower metrics. This demonstrates that misalignment between modalities indeed affects the learning of inter-modality prior, as pointed out in Section \ref{of}, leading to $\text{ProxNet}_Z$'s insufficient exploration of the correlations between different modalities. Furthermore, the zoomed-in regions of interest shows that spatial alignment of corresponding points across modalities can further enable the network to more accurately capture structural information. For instance, as indicated by the arrows in the figure, DUN-SA can recover more clear details. (3) Compared to DUN-SA, w/o DB also exhibits inferior quantitative and qualitative reconstruction performance, indicating the effectiveness of intra-modality prior learning in DUN-SA.

Overall, each key component can help the model enhance reconstruction performance, and the optimal model can be obtained by combining all the modules.

\subsection{Model verification}
\label{ekc}
In this section, we utilize DUN-SA to conduct a model verification experiment to demonstrate the operational mechanism behind the network modules. Fig. \ref{fig14} shows the reconstructed image $x^t$, warped reference image $\mathcal{T}(x_{\text{ref}}, \phi^{t})$ denoted as $x_{ref}^{t}$, inter-modality prior $z^t$ and intra-modality prior $s^t$ at different stages. It can be easily observed that as the number of stages \(t\) increases, the reference image gradually aligns with the fully sampled target image. The results verify the design of our optimization-inspired iterative learning framework—the mutual improvement of the spatial alignment task and the reconstruction task enables the proposed DUN-SA to achieve multi-modal MRI reconstruction as expected. Through these visualization results, the underlying rationale and insights of the proposed network can be intuitively understood, and our network demonstrates better transparency compared to other methods.

\subsection{Computational Complexity Comparison}
For the compared Multi-Modal MRI reconstruction methods (i.e., SAN, MC-CDic, and DUN-SA), Table \ref{tb7} lists two important measures of the complexity, including the inference time and memory requirement for reconstruction of a slice with size 1 $\times$ 320 $\times$ 320 pixels and a 3D volume with size 16 $\times$ 320 $\times$ 320 pixels on an NVIDIA GeForce GTX 3090 GPU. As compared with other methods, DUN-SA exhibits mid-level complexity. However, considering the overall reconstruction performance, DUN-SA presents a favorable cost-performance trade-off.
\begin{table}[!ht]
\centering
\renewcommand{\arraystretch}{0.89}
\small
\setlength\tabcolsep{5pt}
\caption{Complexity analysis of representative models.}
\label{tb7}
\begin{tabular}{ccccc}
\toprule
\textbf{Size} & \textbf{Complexity}  & \textbf{SAN} & \textbf{MC-CDic} & \textbf{DUN-SA} \\
\midrule
\multirow{2}{*}{\shortstack{1 Slice \\ 320 $\times$ 320}}  
                             & Time (s)   & 0.50  & 2.31  & 1.09  \\
                             & Memory (MiB)  & 241.64 & 1089.90 & 398.66 \\
\midrule
\multirow{2}{*}{\shortstack{16 Slices \\ 320 $\times$ 320}} 
                             & Time (s)   & 2.51  & 33.02  & 7.02  \\
                             & Memory (MiB)  & 1338.79 & 16674.56 & 1554.41 \\
\bottomrule
\end{tabular}
\end{table}

\section{Conclusion}
In this paper, we propose a novel joint alignment and reconstruction model for multi-modal MRI reconstruction. By developing an aligned cross-modal prior term, we integrate the spatial alignment task into the reconstruction process. We design an optimization algorithm for solving it and then unfold each iterative stage into the corresponding network module. As a result, we have constructed a deep unfolding network with interpretability, termed DUN-SA. Through end-to-end training, we fully leverage both intra-modality and inter-modality priors. Comprehensive experiments conducted on four real datasets have demonstrated that the proposed DUN-SA outperforms current state-of-the-art methods in both quantitative and qualitative assessments. Additionally, we have verified that DUN-SA is relatively robust to misalignment, with minimal impact on spatial alignment even as acceleration factors increase.

\section*{Acknowledgments}
This research is supported by the National Key R \& D Program of China (No. 2021YFA1003004) and the National Natural Science Foundation of China (No. 11971296).

\bibliographystyle{model2-names.bst}\biboptions{authoryear}
\bibliography{refs}

\end{document}